\documentclass[twocolumn]{aastex63}
\usepackage{booktabs}

\newcommand{\nh} {$N_{\text{H}}$}
\newcommand{\chisq} {$\chi^{2}$}

\newcommand{\nustar} {\textit{NuSTAR}}
\newcommand{\xmm} {\textit{XMM-Newton}}

\newcommand {\msun} {M$_{\odot}$}
\newcommand{\degree} {$^\circ$}

\shorttitle{Modeling disk precession in Her X-1}
\shortauthors{Brumback et al.}

\begin{document}

\title{A broad-band X-ray view of the precessing accretion disk and pre-eclipse dip in the pulsar Her X-1 with \textit{NuSTAR} and \textit{XMM-Newton}}

\author{McKinley C.\ Brumback}
\affiliation{Department of Physics \& Astronomy, Dartmouth College, 6127 Wilder Laboratory, Hanover, NH 03755, USA}
\affiliation{Cahill Center for Astronomy and Astrophysics, California Institute of Technology, 1216 E California Blvd, Pasadena, CA 91125, USA}

\author{Ryan C.\ Hickox}
\affiliation{Department of Physics \& Astronomy, Dartmouth College, 6127 Wilder Laboratory, Hanover, NH 03755, USA}

\author{Felix S.\ F\"urst}
\affiliation{European Space Astronomy Centre (ESA/ESAC), Operations Department, Villanueva de la Ca$\tilde{\text{n}}$ada Madrid, Spain}

\author{Katja Pottschmidt}
\affiliation{CRESST, Department of Physics and Center for Space Science and Technology, UMBC, Baltimore, MD 210250, USA}
\affiliation{NASA Goddard Space Flight Center, Code 661, Greenbelt, MD 20771, USA}

\author{John A.\ Tomsick}
\affiliation{Space Sciences Laboratory, University of California, Berkeley, 7 Gauss Way, Berkeley, CA 94720, USA}

\author{J\"orn Wilms}
\affiliation{Dr. Karl Remeis-Sternwarte and Erlangen Centre for Astroparticle Physics, Sternwartstrasse 7, 96049 Bamberg, Germany}

\author{R\"udiger Staubert}
\affiliation{Institut f\"ur Astronomie und Astrophysik, Universit\"at T\"ubingen, Sand 1, 72076 T\"ubingen, Germany}

\author{Saeqa Vrtilek}
\affiliation{Harvard-Smithsonian Center for Astrophysics, 60 Garden Street, Cambridge, MA 02138, USA}

\submitjournal{The Astrophysical Journal}
\accepted{27 Jan 2021}
\received{7 Aug 2020}
\revised{15 Dec 2020, 26 Jan 2021}

\begin{abstract}
We present a broad-band X-ray timing study of the variations in pulse behavior with superorbital cycle in the low-mass X-ray binary Her X-1. This source shows a 35-day superorbital modulation in X-ray flux that is likely caused by occultation by a warped, precessing accretion disk. Our data set consists of four joint \textit{XMM-Newton} and \textit{NuSTAR} observations of Her X-1 which sample a complete superorbital cycle. We focus our analysis on the first and fourth observations, which occur during the bright ``main-on" phase, because these observations have strongly detected pulsations. We added an archival \textit{XMM-Newton} observation during the ``short-on" phase of the superorbital cycle since our observations at that phase are lower in signal to noise. We find that the energy-resolved pulse profiles show the same shape at similar superorbital phases and the profiles are consistent with expectations from a precessing disk. We demonstrate that a simple precessing accretion disk model is sufficient to reproduce the observed pulse profiles. The results of this model suggest that the similarities in the observed pulse profiles are due to reprocessing by a precessing disk that has returned to its original precession phase. We determine that the broad-band spectrum is well fit by an absorbed power law with a soft blackbody component, and show that the spectral continuum also exhibits dependence on the superorbital cycle. We also present a brief analysis of the energy resolved light curves of a pre-eclipse dip, which shows soft X-ray absorption and hard X-ray variability during the dip.
\end{abstract}

\section{Introduction} \label{Hdisksec:int}
Magnetically dominated accretion occurs regularly in the environments surrounding neutron stars that accrete from a stellar companion via mass transfer mechanisms such as stellar outflows and Roche lobe overflow (e.g., \citealt{nagase2001}). Near the pulsar's magnetosphere, magnetic pressure from the strong magnetic field overwhelms the ram pressure from the accretion disk, which causes the hot, ionized gas to accrete along the dipolar field lines. The structure of the accretion disk and magnetized accretion flow are thought to be complex in nature (e.g., \citealt{ogilvie2001,romanova2002,romanova2003,romanova2004}), however observational constraints on magnetically dominated accretion structures are limited.

The luminous X-ray pulsars LMC X-4, SMC X-1, and Her X-1 show superorbital periods, which are variations on time scales longer than the orbital period. These superorbital periods are attributed to warped disks, where radiation pressure drives a tilt in the accretion disk (e.g., \citealt{pringle1996,ogilvie2001}). As the disk precesses, the warped edge temporarily occults the pulsar and causes the changes in luminosity. The geometry of the accretion disk and the neutron star in these warped disk systems creates some of the best observational laboratories for studying magnetized accretion flow onto compact objects because emission from the accretion and reprocessed emission from the accretion disk can both be detected (\citealt{hickox2004}). \cite{brumback2020} (hereafter B20) used pulse phase-resolved spectroscopy and tomography to model the warped inner accretion disk structure around the bright X-ray pulsars LMC X-4 and SMC X-1. This analysis assumed that the supeorbital period, which is present on timescales of 30-60 days in these sources, is caused by the precession of a warped inner accretion disk seen approximately edge-on (e.g., \citealt{gerend1976,heemskerk1989,wojdowski1998}). In sources with this accretion disk geometry, the rotating, high energy beam from the neutron star irradiates the inner accretion disk. The disk reprocesses the beam's radiation and releases softer X-rays which differ in pulsation shape and phase from the hard pulses (\citealt{shulman1975,neilsen2004,zane2004,hickoxvrtilek2005}). B20 used the simple geometric warped disk model created by \cite{hickoxvrtilek2005} to model observed changes in hard and soft pulse profile shape as a function of superorbital phase in LMC X-4 and SMC X-4 with broadband X-ray coverage.

In this work, we will follow the method of B20 and provide observational constraints on the geometry of the warped inner accretion disk in Her X-1, a prototype X-ray pulsar (\citealt{tananbaum1972}), making use of full hard X-ray coverage from the {\em Nuclear Spectroscopic Telescope Array} ({\em NuSTAR}). Her X-1 is a low mass X-ray binary consisting of a 1.5 \msun\ neutron star orbiting HZ Herculis, an approximately 2.2 \msun\ A/F type star (\citealt{crampton1974,deeter1981,reynolds1997}). The binary orbit is 1.7 days and is nearly circular (\citealt{staubert2009}). The orbital plane is highly inclined ($i=85-88$\degree), resulting in regular eclipses of the neutron star. The distance to the binary is 6.6 kpc (\citealt{reynolds1997}). Her X-1 is a cyclotron line source (\citealt{truemper1978}), with a broad absorption feature caused by cyclotron resonance scattering being visible in the hard X-ray spectrum. The energy at which this feature occurs is caused by the strength of the pulsar's magnetic field, and thus the feature allows astronomers to directly measure the field strength. The central energy of this cyclotron resonance scattering feature (CRSF) has been found to correlate strongly with the source flux (\citealt{staubert2007,staubert2016,staubert2017}). In addition, the energy shows an interesting evolution with time: after a fairly constant value around 37 keV for about a decade after the discovery, a strong turn-up occurred to beyond 40 keV (\citealt{gruber2001}), followed by a nearly 20 yr decline until $\sim$2012, after which the line energy is again constant around 37 keV (\citealt{staubert2016,staubert2020,bala2020}). This value suggests that the magnetic field in Her X-1 is approximately $3.2\times10^{12}$ G (\citealt{staubert2020}). \cite{tananbaum1972} discovered 1.24 s X-ray pulsations from the neutron star and a longer, approximately 35 day modulation in X-ray luminosity.

This 34.85 day superorbital period in Her X-1 is likely caused by a warped, precessing inner accretion disk (\citealt{giacconi1973, ramsay2002,zane2004}). The superorbital period consists of a bright ``main on" period lasting approximately 11 days and a shorter, fainter ``short on" period lasting approximately 8 days, which are separated by 8 day off states (\citealt{tananbaum1972,giacconi1973}). During the bright main on state, Her X-1 reaches characteristic X-ray luminosities of approximately $10^{37}$ erg s$^{-1}$. During the 35 day cycle, the neutron star eclipses every 1.7 days and there are additional pre-eclipse dips seen every 1.62 days (\citealt{jones1976}). These pre-eclipse dips are likely caused by obscuration of the pulsar by material at the interaction point between the accretion stream and the disk (e.g., \citealt{still2001}). Another feature of the superorbital behavior in Her X-1 are the anomolous low states (ALSs), which are defined by a significant decrease in X-ray flux and pulsation strength without significant change in the optical and UV flux (e.g., \citealt{vrtilek1996,coburn2000,staubert2017}). These states can last for several months and are likely caused by a change in scale height of the accretion disk, which obscurs the central accretor, or a change in disk inclination (e.g., \citealt{parmar1985}).

In this work we present joint \xmm\ and \nustar\ observations of Her X-1 that sample a single superorbital cycle, allowing us to monitor and model the precession of the accretion disk. Although the source is well-studied, previous Her X-1 observations lack the combination of energy coverage, timing resolution, and sampling of different phases within a single superorbital cycle needed to fully constrain the pulsar beam and disk geometry over the precession cycle of the disk. \cite{ramsay2002} and \cite{zane2004} used \xmm\ to observe the precession of the inner accretion disk, but their analysis lacked \nustar\ high energy coverage. \cite{fuerst2013} illustrated the importance of high energy pulse profiles in Her X-1 by showing the dramatic changes in pulse shape and phase that occur across the \nustar\ energy band. However, the \cite{fuerst2013} observations took place during one superorbital phase so they could not demonstrate changes in pulse-profile at different superorbital phases. \cite{mccray1982} and \cite{vrtilek1985} had soft X-ray coverage of a complete 35 day cycle of Her X-1 with \textit{Einstein}'s MPC. However, \textit{Einstein}'s energy range of 1--20 keV does not fully constrain changes in the soft pulsations, which \cite{hickox2004} found peak below 1 keV. \cite{staubert2013} demonstrated the  variability of the pulse profiles of Her X-1 as a function of superorbital phase, their observations do not have the soft X-ray coverage necessary to capture reprocessed emission (less than 1 keV) from the accretion disk.

\begin{figure*}
\centering
\includegraphics[scale=0.9]{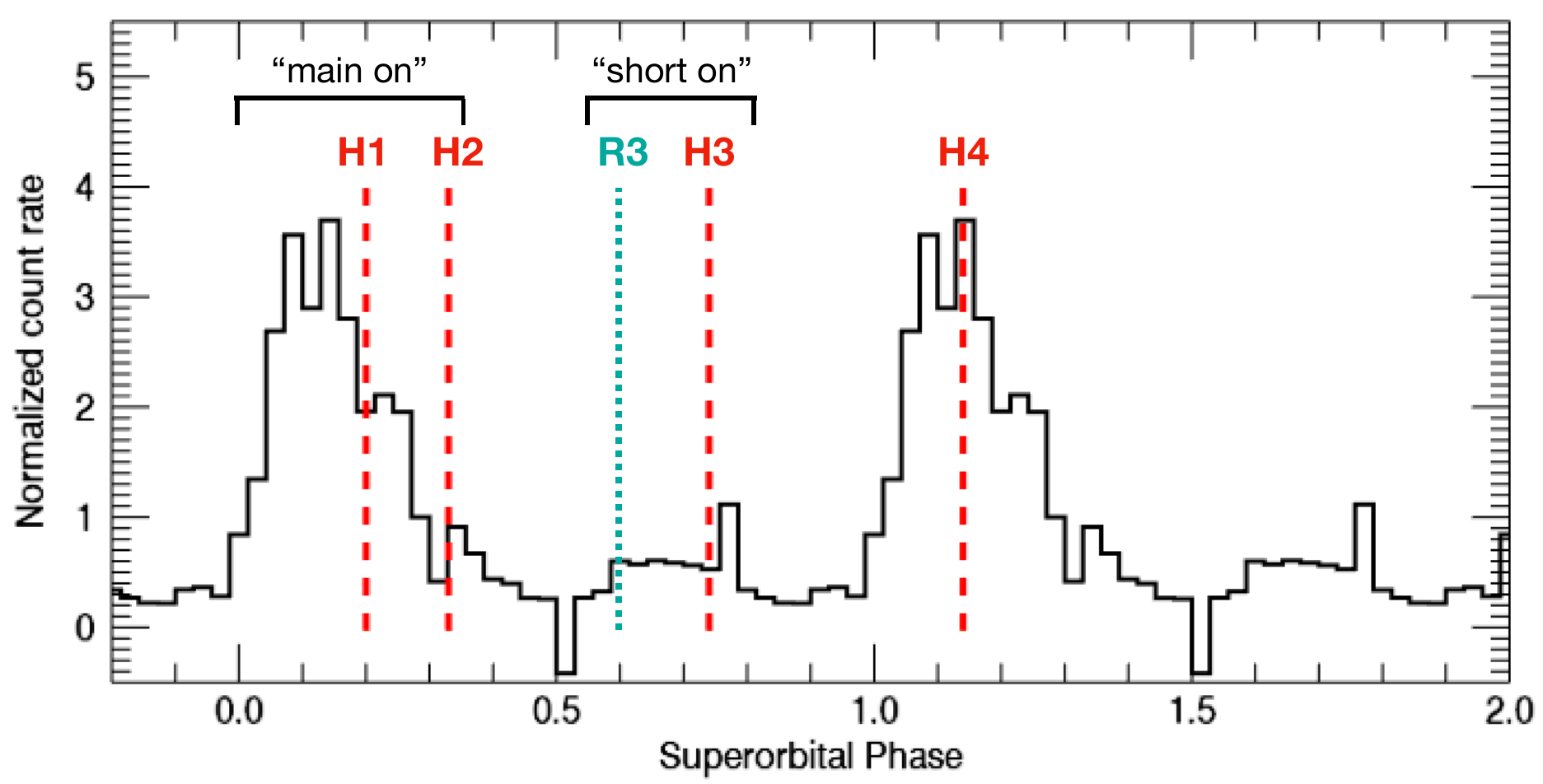}
\caption{The one day averaged 2--20 keV MAXI (\citealt{matsuoka2009}) light curve of Her X-1, which we folded on the 35 day superorbital cycle and plotted twice for clarity. We show the times of the joint \xmm\ and \nustar\ observations used in this analysis as red vertical, dashed lines. We label the observations H1 through H4 to show their place in this series. The teal vertical dotted line shows the superorbital phase of the archival \xmm\ observation 0111061201; this is the third observation in a series of observations first analyzed by \cite{ramsay2002}, which we refer to as Observation R3. The main on and short on time periods of the superorbital cycle are marked with brackets.}
\label{fig:hersolc}
\end{figure*}

In this paper we use both \nustar\ and \xmm\ in order to conduct a broad-band study of the pulse profiles and spectral shapes of Her X-1 during two consecutive main-on states of the superorbital cycle, allowing us to test the periodic dependence of these features.  We also supplement our observations with archival observations that allow us to study the short-on state of the superorbital cycle.

In Section \ref{Hdisksec:obs} we describe the previously unpublished joint \xmm\ and \nustar\ observations of Her X-1, as well as some archival data used in this analysis. In Section \ref{Hdisksec:analysis} we describe the procedure used to extract pulse profiles and perform phase-averaged and phase-resolved spectroscopy and we present the results in Section \ref{Hdisksec:results}. In Section \ref{Hdisksec:diskmodel} we model the observed pulse profiles and simulate the inner disk geometry. We briefly examine the energy resolved light curves of a pre-eclipse dip in one of our observations in Section \ref{Hdisksec:dip}. In Section \ref{Hdisksec:disc} we discuss the implication of our results.

\section{Observations} \label{Hdisksec:obs}
The data used in this analysis are a set of four joint \nustar\ (\citealt{harrison2013}) and \xmm\ observations of Her X-1 that took place between 09 February and 14 March 2019. Following the convention used in B20, we refer to these observations as Observation H1, H2, H3, and H4. Table \ref{tab:herdat} shows the date of observation, the observation ID number, superorbital phase, and exposure time for each observation. Figure \ref{fig:hersolc} shows the one day averaged MAXI light curve, folded on the 35 day superorbital period, for Her X-1 with the observations plotted as red vertical lines.

\begin{deluxetable*} {cccccccc}
\label{tab:herdat} 
\tablecolumns{7}
\tablecaption{Description of Her X-1 Observations\tablenotemark{a}}  
\tablewidth{0pt}
\tablehead{
\colhead{Name} & \colhead{Date} & \colhead{Mid-exposure Time (MJD)} & \colhead{$\phi_{SO}$} & \colhead{Observation ID} & \colhead{Observatory} & \colhead{Telescope Mode}
& \colhead{Exposure (ks)}  }
\startdata
H1 & 09 Feb.\ 2019 & 58523.619 & 0.20 & 30402034002 & \nustar\ & \nodata & 18.2 \\
H1 & 09 Feb.\ 2019 & 58523.669 & 0.20 & 0830530101 & \xmm\ & Fast Timing Mode & 21.9 \\
H2 & 13 Feb.\ 2019 & 58528.147 & 0.33 & 30402034004 & \nustar\ & \nodata & 22.0 \\
H2 & 13 Nov.\ 2019 & 58528.131 & 0.33 & 0830530201& \xmm\ & Fast Timing Mode & 27.2 \\
R3\tablenotemark{b} & 16 Mar.\ 2001 & 51984.992 & 0.60 & 0111061201 & \xmm\ & Fast Timing Mode & 11 \\
H3 & 28 Feb.\ 2019 & 58542.515 & 0.74 & 30402034006 & \nustar\ & \nodata & 23.0 \\
H3 & 28 Feb.\ 2019 & 58542.534 & 0.74 & 0830530301 & \xmm\ & Fast Timing Mode & 32.6 \\
H4 & 14 Mar.\ 2019 & 58556.515 & 1.14 & 30402034008 & \nustar\ & \nodata & 16.7 \\
H4 & 14 Mar.\ 2019 & 58556.523 & 1.14 & 0830530401 & \xmm\ & Fast Timing Mode & 29.1
\enddata
\tablenotetext{a}{These observations span two consecutive superorbital cycles. The turn on for the first cycle was 58516.6 MJD and the turn on for the second cycle was 58551.5 MJD. }
\tablenotetext{b}{This is the third observation in an archival data set of Her X-1 first presented in \cite{ramsay2002} and later used in \cite{zane2004} and \cite{jimenesgarate2002}. In this work, we use Observation R3 as an alternative to the weak pulsations in Observation H3.}
\end{deluxetable*}

We reduced these data using HEASoft version 6.26.1 with \nustar\ CALDB v20191219 and XMMSAS version 18.0.0. For the \nustar\ observations, we selected a circular source region of radius 110 arcseconds with DS9 (\citealt{ds9}). We used a circular region of the same size away from the source as the background region. The background counts make up approximately 0.3\% of the total counts. For the \xmm\ observations we used the EPIC-pn instrument in Timing Mode exclusively to obtain the maximum possible timing resolution and to minimize the effects of pileup, and selected only single and double events from the EPIC-pn data. We used the XMMSAS tool {\fontfamily{qcr}\selectfont epatplot} to evaluate the observations for pileup by comparing the model of expected single and double events to data. In the observations that took place during Her X-1's bright main on phase (obsIDs 0830530101 and 0830530401), we found differences in the models indicative of pileup. We excised the brightest central pixels and examined the models again until we found good agreement. Ultimately, we found it was only necessary to exclude the central pixel to reduce pileup.

We performed a barycentic correction to Observations H1--H4 using the NuSTARDAS tool {\fontfamily{qcr}\selectfont barycorr} and the XMMSAS tool {\fontfamily{qcr}\selectfont barycen}. We accounted for the effect of the neutron star's orbit by correcting the photon arrival times using the Her X-1 orbital ephemeris described in \cite{staubert2009}.

\begin{figure*}
\centering
\includegraphics[scale=0.65]{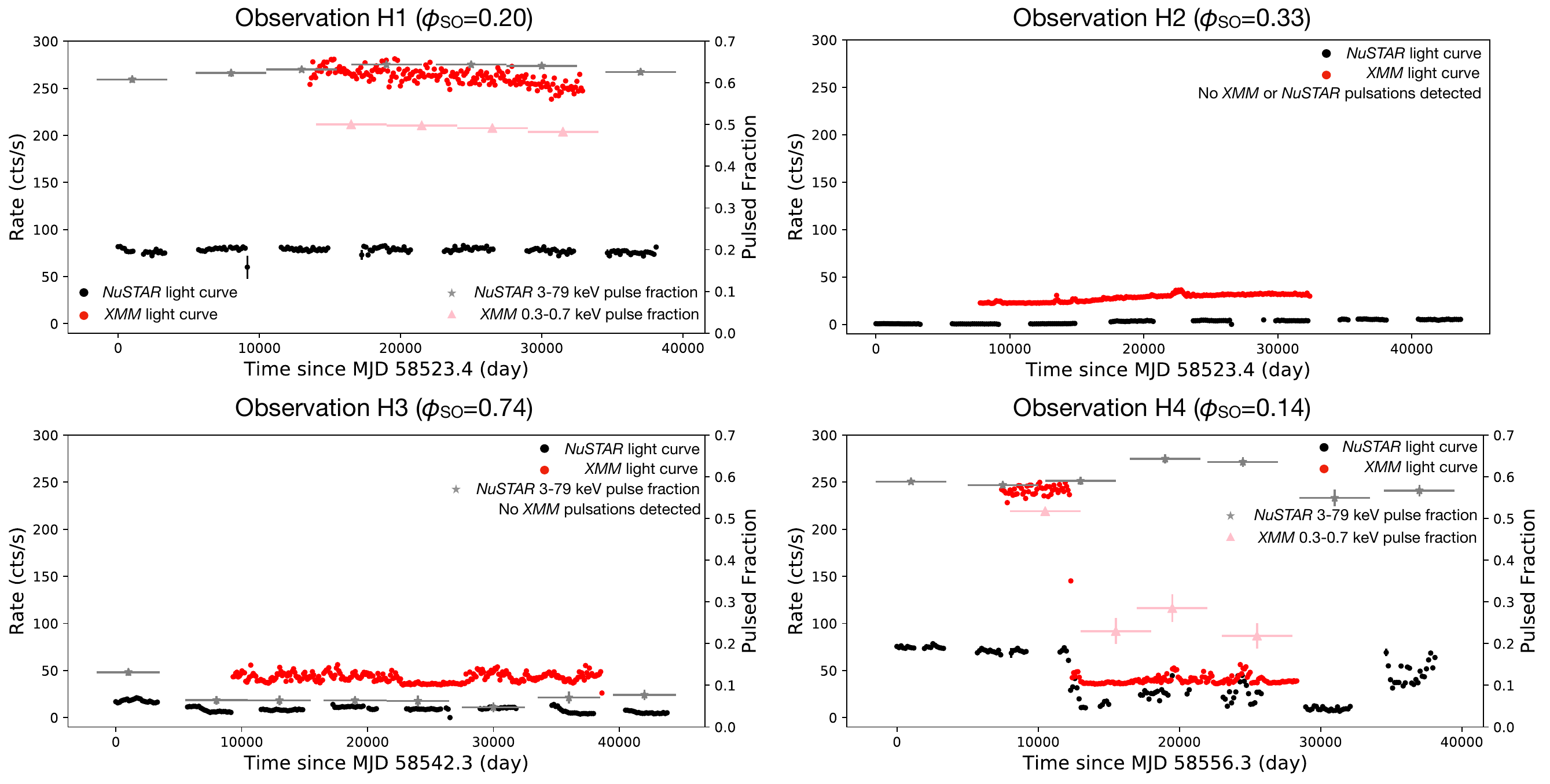}
\caption{\nustar\ 3--79 keV (black) and \xmm\ 0.2--12 keV (red) light curves of the four joint Her X-1 observations. We also show the pulsed fractions, discussed in Section \ref{Hdisksec:analysis}, for the 3--79 keV \nustar\ data (grey stars) and 0.3--0.7 keV \xmm\ data (pink triangles) where pulsations are detected. We offset the \xmm\ light curves by a value of 20 in Observations H2 and H3 and 30 in Observation H4 for clarity. A pre-eclipse dip is visible in both the \nustar\ and \xmm\ light curves of Observation H4. For this analysis, we only used data where pulsations were strongly detected in both \xmm\ and \nustar. For clarity, we have filtered the light curves of the bright observations H1 and H4 in order to remove bins with low exposure fraction caused by \xmm's Counting Mode.}
\label{fig:herlc}
\end{figure*}

We show the \nustar\ 3--79 keV and \xmm\ 0.2-15 keV light curves for the Her X-1 observations in Figure \ref{fig:herlc}. The \xmm\ count rates have been arbitrarily offset from the \nustar\ count rates in Observations H2, H3, and H4 for clarity. Observation H4 captured a pre-eclipse dip with both telescopes.
We also show the pulsed fractions for the \nustar\ 3--79 keV data and the \xmm\ 0.3--0.7 keV data overplotted with the light curves. Plotting the hard and soft pulsed fractions allowed us to determine which parts of the data set were suitable for our analysis, which required both hard and soft pulsations (see Section \ref{Hdisksec:diskmodel}). For this reason, we did not include Observation H2 (where pulsations were not detected in \xmm\ or \nustar), Observation H3 (where \xmm\ pulsations were not detected), or data after the onset of the dip in H4 (where the soft pulses dramatically weaken).

Because we were unable to extract both hard and soft pulse profiles from Observations H2 and H3, we turned to archival data of Her X-1 to increase our coverage of the 35 day superorbital cycle. As mentioned in the introduction, many previous works have presented changes in pulse profile shape across the superorbital cycle. We focused on the \xmm\ data archive because the 0.2--12 keV energy range would allow us to directly compare the soft data to our observations while offering some overlap with our \nustar\ data. For the sake of brevity, we selected a single archival \xmm\ observation of Her X-1. ObsID 0111061201 took place on 16 March 2001 as part of a series of \xmm\ observations of Her X-1 and was first presented in \cite{ramsay2002}. Because this observation is the third in its original series, we refer to it as Observation R3 in this work. Observation R3 took place at superorbital phase 0.60. We reduced these data following the analysis steps described in \cite{ramsay2002}, which are broadly consistent with the reduction process described in this work. However, we added the correction described in \cite{ramsay2002} to {\fontfamily{qcr}\selectfont FTCOARSE}, which had erroneously incremented by 1 second during this observation. As in \cite{ramsay2002}, we only examine the EPIC-pn data to minimize pile up. We used the XMMSAS tool {\fontfamily{qcr}\selectfont barycen} to apply a barycentric correction to this data set. However we did not use the \cite{still2001} orbital ephemeris for Her X-1, which was used by \cite{ramsay2002}, and instead chose to correct the photon arrival times with the more recently updated \cite{staubert2009} ephemeris.

\section{Data Analysis} \label{Hdisksec:analysis}

\subsection{Timing Analysis}
We used HENDRICS's $Z^{2}_{4}$ statistics search, which is functionally similar to epoch folding, to determine the spin period and spin period derivative of each observation (\citealt{buccheri1983,hendrics}). For all observations used in this work, we found that the spin period derivative was consistent with zero. We estimated the uncertainty in the spin period by creating pulse profiles from the beginning and end of each observation and calculating $\delta_{\text{frequency}} = \delta_{\text{phase}} / \delta_{\text{time}}$. While we were able to precisely measure the spin period in Observations H1 and H4 due to their length and good signal to noise ratio, the fainter observations (which do not show pulsations across the entire observations) had higher uncertainties, as can be seen in Table \ref{tab:hperiod}. We do not list any period for Observation H2 because we were unable to detect pulsations during this observation. All errors are 90\% confidence unless otherwise specified.

For the archival data set, Observation R3, our use of a different orbital ephemeris necessitated redoing the pulsation search, which we did using the method described above. We detected pulsations with a spin period of 1.2379 $\pm$ 0.0001 s, which is slightly larger than the best period found by \cite{ramsay2002}. This difference is most likely caused by the use of a different orbital ephemeris. To check for consistency with \cite{ramsay2002}, we extracted energy resolved pulse profiles in a soft band of 0.3--0.7 keV and a hard band of 8--12 keV. We found good agreement between our energy resolved pulse profiles and those presented in Figure 2 of \cite{ramsay2002}.

\begin{deluxetable} {cc} 
\label{tab:hperiod}
\tablecolumns{2}
\tablecaption{Best Fit Spin Periods for Her X-1 Observations}  
\tablewidth{0pt}
\tablehead{
\colhead{Observation} & \colhead{Spin Period (s)\tablenotemark{a}} }
\startdata
H1 & 1.237721(6) \\
H2 & N/A \\
H3  & 1.2377(4) \\
R3 & 1.2379(2) \\
H4 & 1.237721(8) 
\enddata
\tablenotetext{a}{For Observations H1--H4, we used \nustar\ 3--79 keV data to determine the pulse period. For Observation R3, we used \xmm\ EPIC-pn 0.2--12 keV data.}
\end{deluxetable}

We filtered the Observations H1 and H4 data sets by energy, selecting the 8--60 keV energy range from the \nustar\ data and the 0.3--0.7 keV energy range from the \xmm\ data. We selected these energy ranges specifically for two reasons. The first is to separate soft, reprocessed emission (less than 1 keV) from the accretion disk and hard emission from the pulsar beam. The second was that an analysis of the energy-resolved pulse profiles indicated changes in the soft pulse profile beginning at approximately 0.8 keV. We show the energy resolved pulse profiles and discuss those results in more detail in Appendix A.

\begin{figure*}
\centering
\begin{tabular}{cc}
    \includegraphics[scale=0.6]{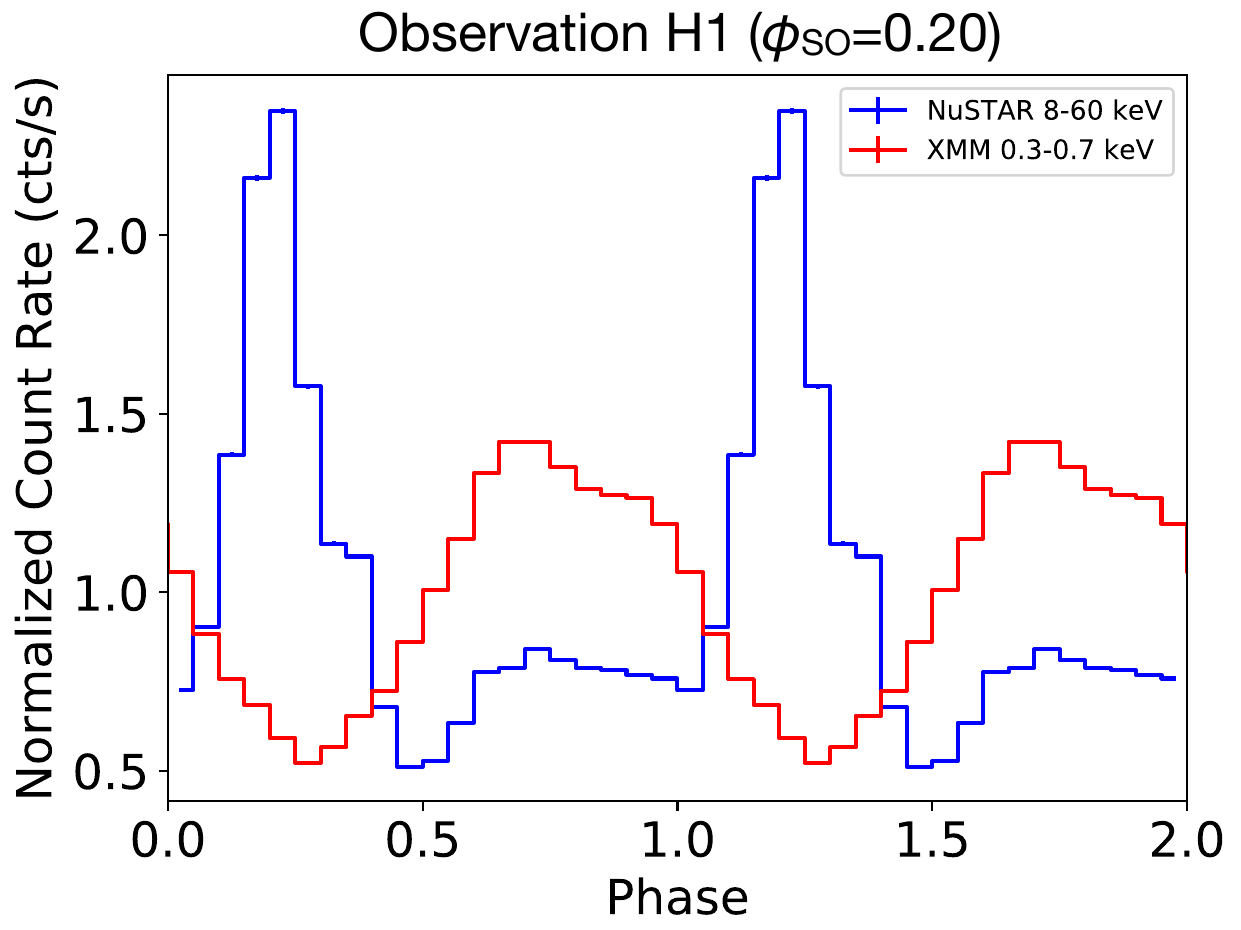} & \includegraphics[scale=0.6]{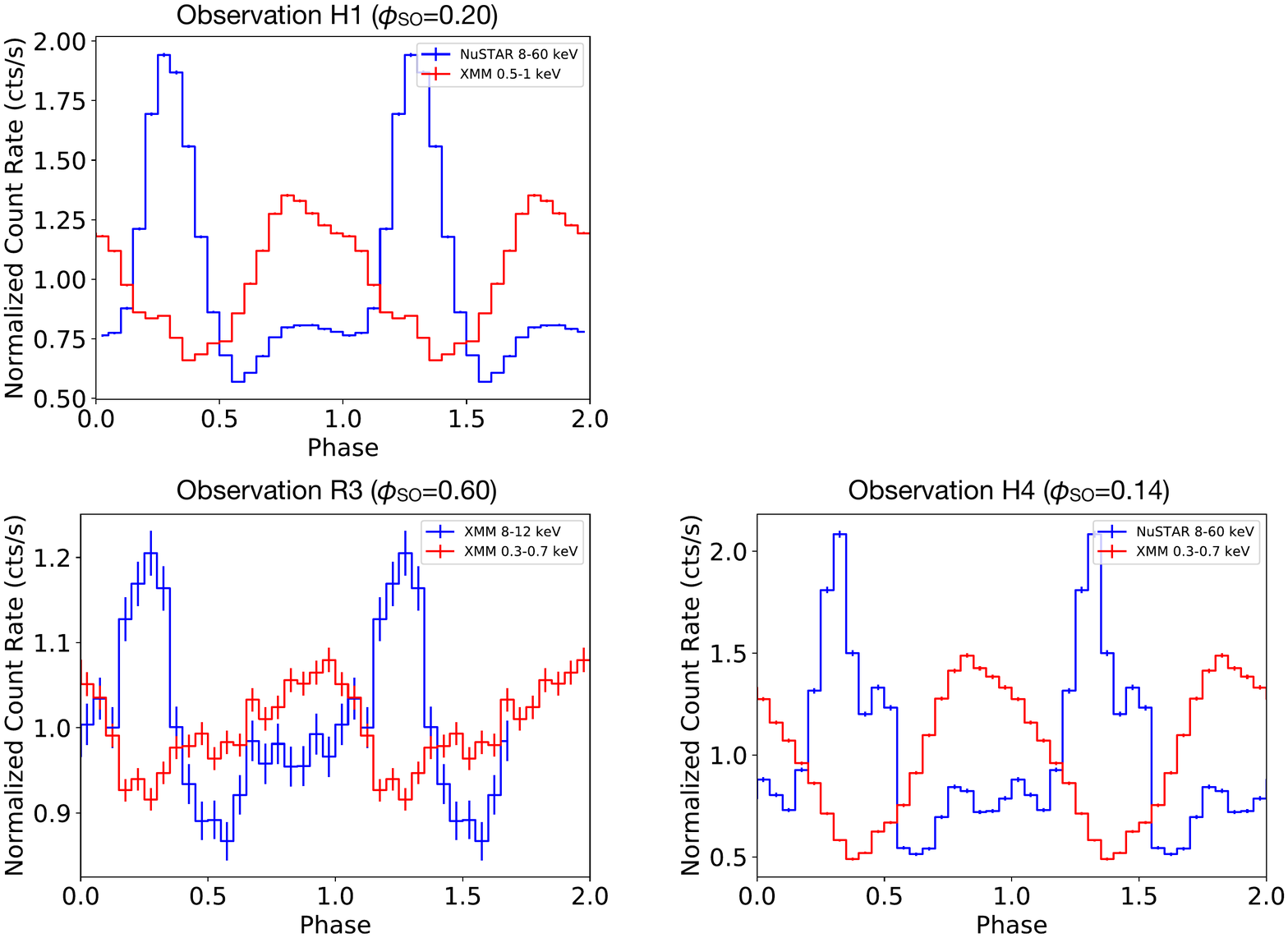} \\
    \multicolumn{2}{c}{\includegraphics[scale=0.6]{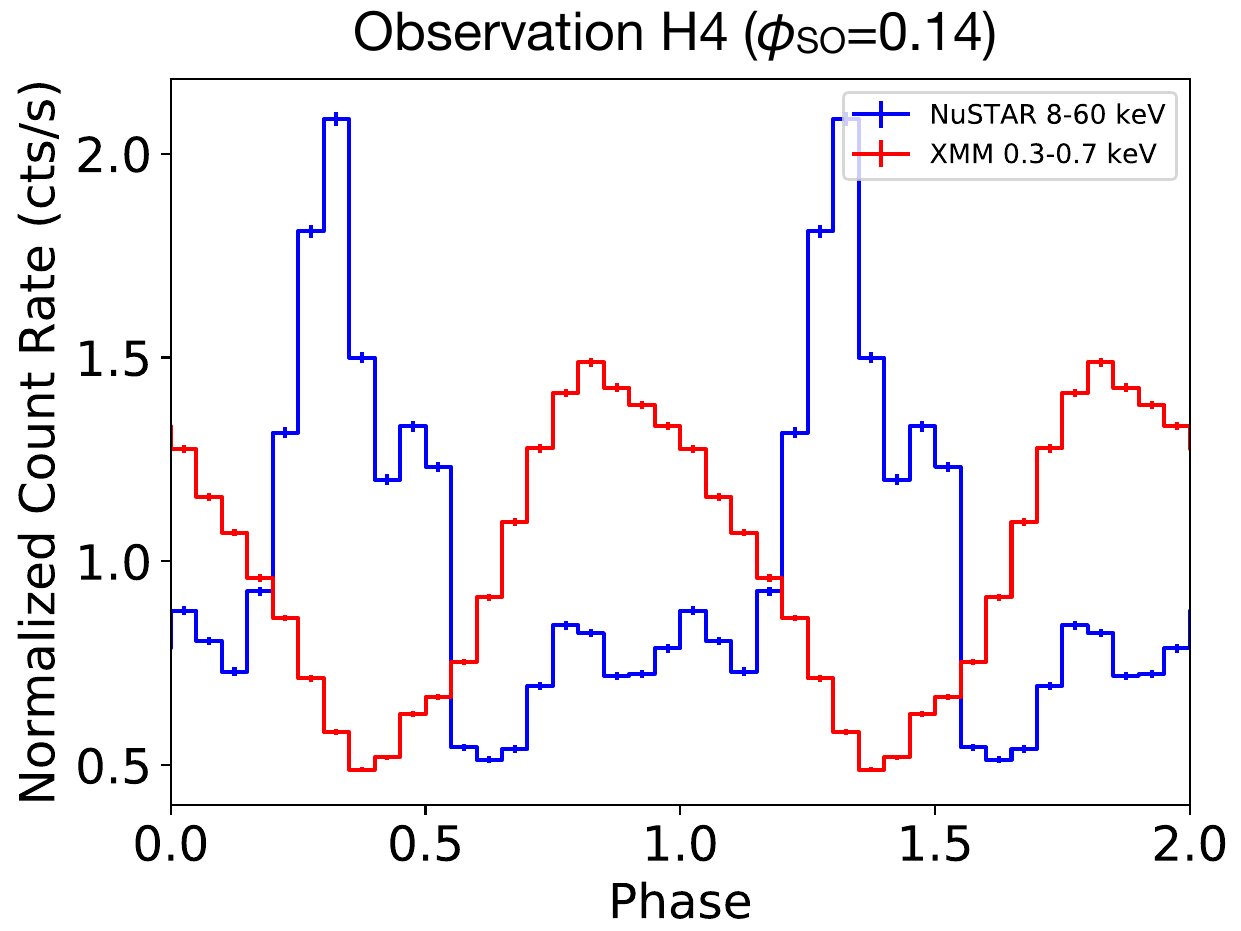}}
\end{tabular}
\caption{Joint pulse profiles for the Her X-1 observations H1, R3, and H4. The same relative phase shift between the \nustar\ 8--60 keV pulse profile (blue) and the \xmm\ 0.3--0.7 keV pulse profile (red) is visible in Observations H1 and H4, which we expected because these observations occur at similar superorbital phases. However the soft pulse profile shown in Observation R3 shows a change in shape and relative phase because of its different superorbital phase. To highlight these relative phase shifts, we shifted the pulse profiles of R3 and H4 so that the hard peaks were aligned with those of Observation H1.} The slight difference between superorbital phase 0.20 (Observation H1) and 0.14 (Observation H4) can be seen in the subtle differences in shape of the \nustar\ pulse profile.
\label{fig:herpp}
\end{figure*}

\begin{figure*}
    \centering
    \includegraphics[scale=0.8]{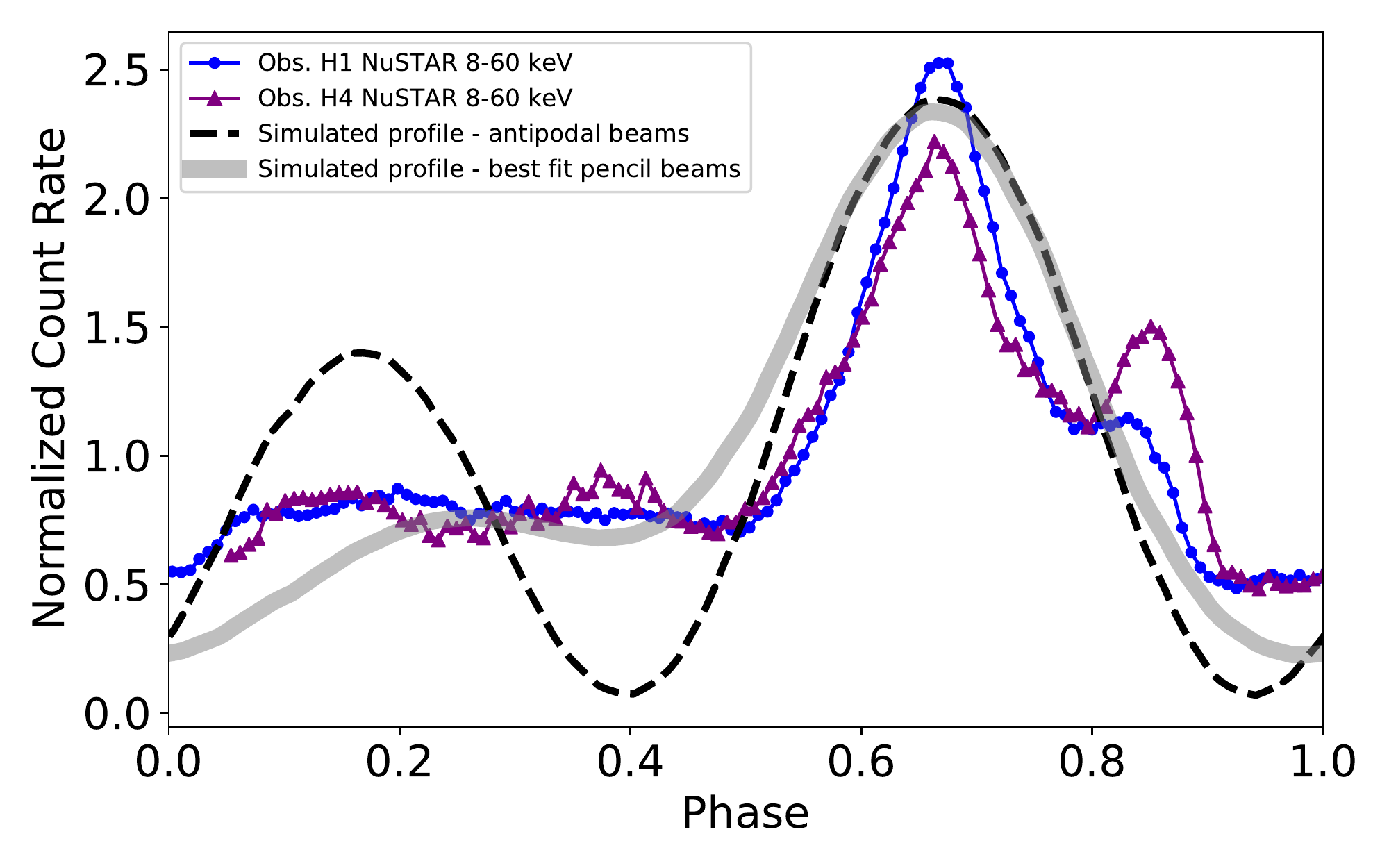}
    \caption{Pulse profiles in the hard \nustar\ band for Observation H1 (blue circles) and Observation H4 (purple triangles) binned with 128 bins per phase. Fine structure in the hard pulse profiles are visible including two bumps within the interpulse of Observation H4 and a notch in the main pulse of both observations. The differences in fine structure between Observations H1 and H4 originate from the small difference in superorbital phase between these observations, which emphasizes that the pulse profiles are excellent trackers of superorbital phase (e.g., \citealt{fuerst2013,staubert2013}). We also show the simulated pulse profiles from an antipodal pencil beam configuration (black dashed line) and our best fit non-antipodal pencil beam (grey line) from our warped disk model (see Section \ref{Hdisksec:diskmodel}). Both of these simulated pulse profiles were binned with 128 bins per phase and smoothed with a Savitzky-Golay filter. Our warped disk model is not capable of reproducing fine structure in the pulse profiles, and we therefore fit the coarser structure. We can also see that an antipodal beam geometry is insufficient to describe the hard pulse shape in Her X-1 with our model.}
    \label{fig:highrespp}
\end{figure*}

We then used the epoch folding tool {\fontfamily{qcr}\selectfont fold\_events} from the Stingray (\citealt{stingray}) software to create energy resolved pulse profiles (see Figure \ref{fig:herpp}). The pulse profiles in Figure \ref{fig:herpp} contain 20 bins per pulse phase. We selected this binning based on the resolution of our simulated pulse profiles, which are not capable of reproducing fine structure within the pulse profile (see Figure \ref{fig:highrespp} and discussion in Sections \ref{Hdisksec:results} and \ref{Hdisksec:diskmodel}). When making the pulse profiles, we used the start time of each observation as phase zero for that pulse profile because we are interest in the relative change in phase between hard and soft profiles rather than phase difference between observations. However, to highlight these relative phase shifts, we shifted the pulse profiles of R3 and H4 so that the hard peaks were aligned with those of Observation H1.

Using the above method, we were also able to create a pulse profile from the \nustar\ data of Observation H3. While we could not use this observation in our modeling analysis due to the lack of detected \xmm\ pulsations, we show the \nustar\ pulse profile in Appendix A for completeness.

For data where pulsations were detected, we calculated the pulsed fractions by first dividing the \xmm\ and \nustar\ data into 5 ks time intervals. For each of these time intervals, we calculated the pulsed fractions that we showed in Figure \ref{fig:herlc} as $PF = (P_{\text{max}} - P_{\text{min}}) / (P_{\text{max}} + P_{\text{min}})$, where $P_{\text{max}}$ is the maximum of the pulse profile and $P_{\text{min}}$ is the minimum of the profile. The errors were calculated from a distribution of 100 pulse profiles made with randomly selected periods between 0.5--1 seconds and 1.25--2 seconds (that is, close to but not precisely the actual pulse period).

\subsection{Phase-averaged Spectroscopy}

To create joint spectra we extracted \nustar\ and \xmm\ source and background spectra from the regions described in Section \ref{Hdisksec:obs} using NuSTARDAS and XMMSAS, respectively. However, we did not extract background spectra from the \xmm\ data because the high source flux dominated the EPIC-pn CCD in Timing Mode, leaving no source-free region for background calculation (e.g., \citealt{ng2010}). We grouped the \nustar\ spectra into bins with a signal to noise ratio of 10 and the \xmm\ spectra with 100 counts per bin. In order to fit the spectra over our desired energy range of 0.3--60 keV, for consistency with our timing analysis, we were required to use \xmm\ data outside of the nominal calibration range of 0.7--12 keV for EPIC-pn in Timing Mode. We do not believe this affected the quality of our spectral fits since the features at low energies (the blackbody component and 1 keV bump feature, see below) are strongly preferred by the data and have been seen before in this source (e.g., \citealt{jimenesgarate2002,hickox2004}).

We fitted the phase-averaged spectrum over the 0.3--60 keV energy range using Xspec version 12.10.1 (\citealt{arnaud1996}). We found that the double power law parameters were degenerate with the absorption models used to describe the cyclotron resonance scattering feature (CRSF) when using the Negative and Positive EXponential (NPEX, e.g., \citealt{mihara1998}) to describe the continuum (as was done B20).  Therefore, we used  a continuum model of a power law with a high energy cut-off ({\fontfamily{qcr}\selectfont powerlaw*highecut}) that did not show degeneracy. This continuum model creates a discontinuity at the cut-off energy that we corrected for by adding a Gaussian absorption feature with its energy tied to the cut-off energy and a free width and depth (e.g., \citealt{coburn2002} and references therein).

Following the spectral model of B20, we also included an absorbing column (tbnew), a blackbody with $kT \sim 0.1$ keV, and several Gaussian emission lines corresponding to Fe K$\alpha$ (6.4 keV) and a $\approx$1 keV ``bump". This continuum bump at 1 keV has been previously observed in Her X-1 with \xmm's RGS instrument (\citealt{jimenesgarate2002}) and with \textit{Suzaku} (\citealt{fuerst2013}). Both \cite{jimenesgarate2002} and \cite{fuerst2013} suggest that this feature is an unresolved complex of lines from Ne and Fe. The \xmm\ EPIC-pn camera does not have the spectral resolution to resolve the complex structure of these features and, therefore, we allowed for a single Gaussian emission line at 0.9 keV that encompassed the ``bump" and described the shape of the spectrum phenomenologically. We allowed the Fe K$\alpha$ line to contain both a broad ($\sigma=0.5$ keV) and a narrow ($\sigma=0.1$ keV) component. The Fe emission at 6.4 keV has been shown in previous works to be associated with near-neutral iron within the inner accretion disk and accretion column (e.g., \citealt{pravdo1978,leahy2000}).

We fixed the absorbing column density to 1.5$\times 10^{20}$ cm$^{-2}$, which we calculated using the HI4PI Map (\citealt{hi4pi2016}) and the HEASARC \nh\ calculator. We set the abundances to those described in  \cite{wilms2000} and the cross sections to those from \cite{verner1996}.

Unlike LMC X-4 and SMC X-1 which were modeled in B20, Her X-1 is a cyclotron line source. To model the cyclotron resonance scattering feature (CRSF) we tested two possible models: the Gaussian absorption model {\fontfamily{qcr}\selectfont gabs} and the cyclotron absorption model {\fontfamily{qcr}\selectfont cyclabs} (e.g., \citealt{mihara1990}). While we found that both models were capable of fitting the CRSF with a similar reduced \chisq, the {\fontfamily{qcr}\selectfont cyclabs} model proved degenerate with the continuum model, resulting in unreasonable values for the CRSF energy and the power law folding energy. Additionally, the {\fontfamily{qcr}\selectfont gabs} model is used by \cite{staubert2020} in their long term monitoring of Her X-1's CRSF, and using {\fontfamily{qcr}\selectfont gabs} in our spectral models provides results that are consistent with their analysis. For these reasons, We used the {\fontfamily{qcr}\selectfont gabs} model in both observations.

\begin{figure*}
\centering
\includegraphics[scale=0.42]{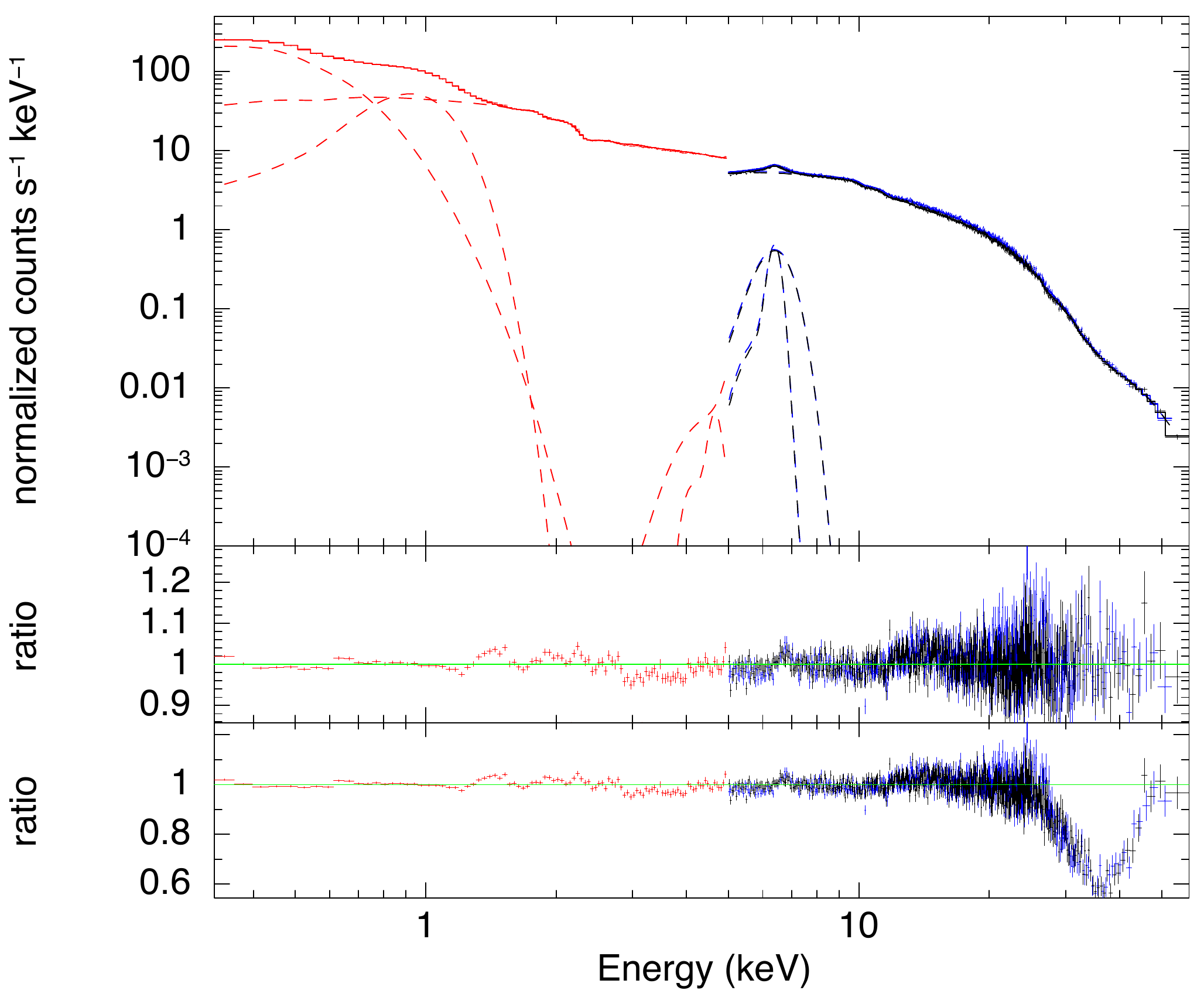} 
\includegraphics[scale=0.42]{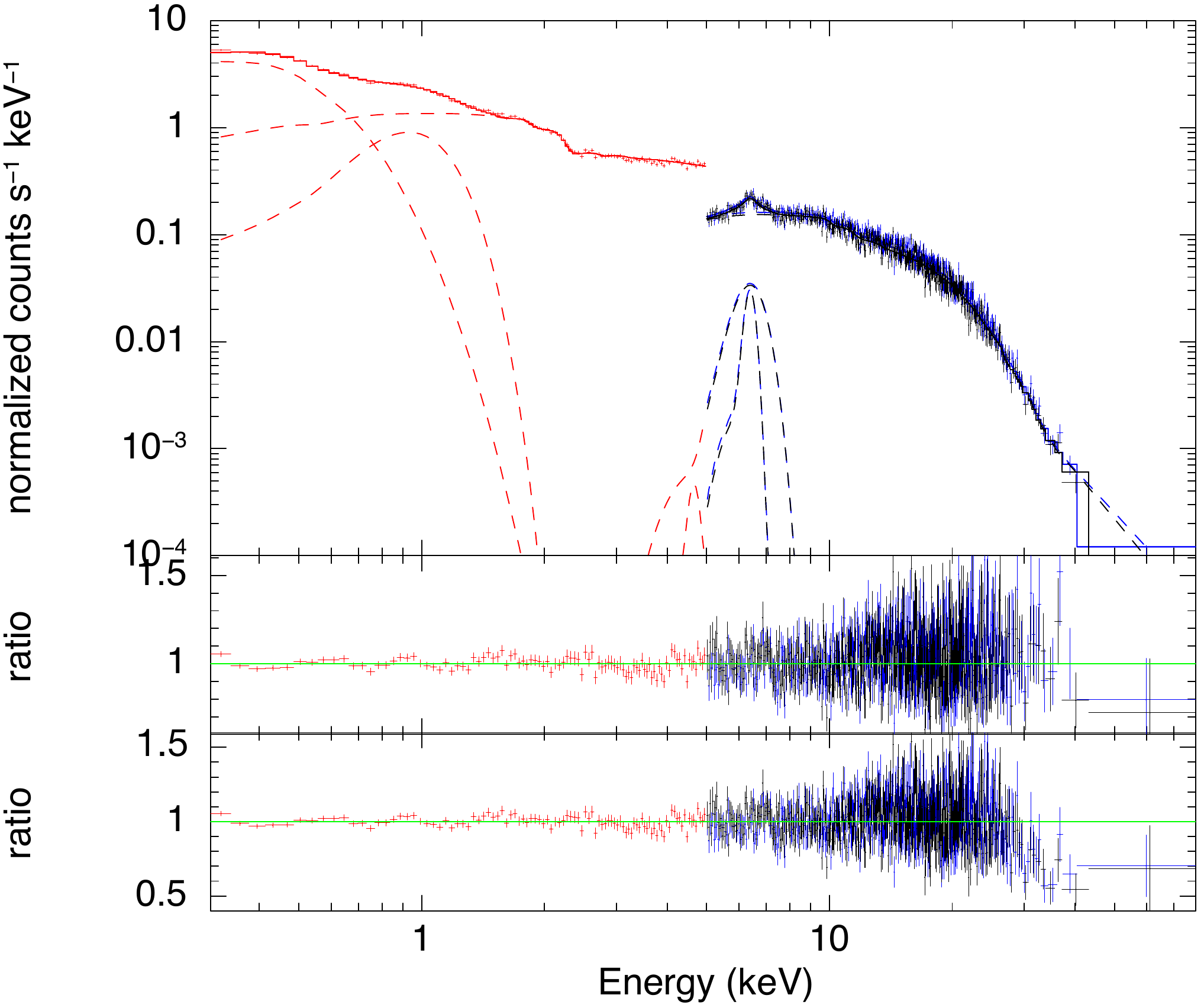} 
\includegraphics[scale=0.42]{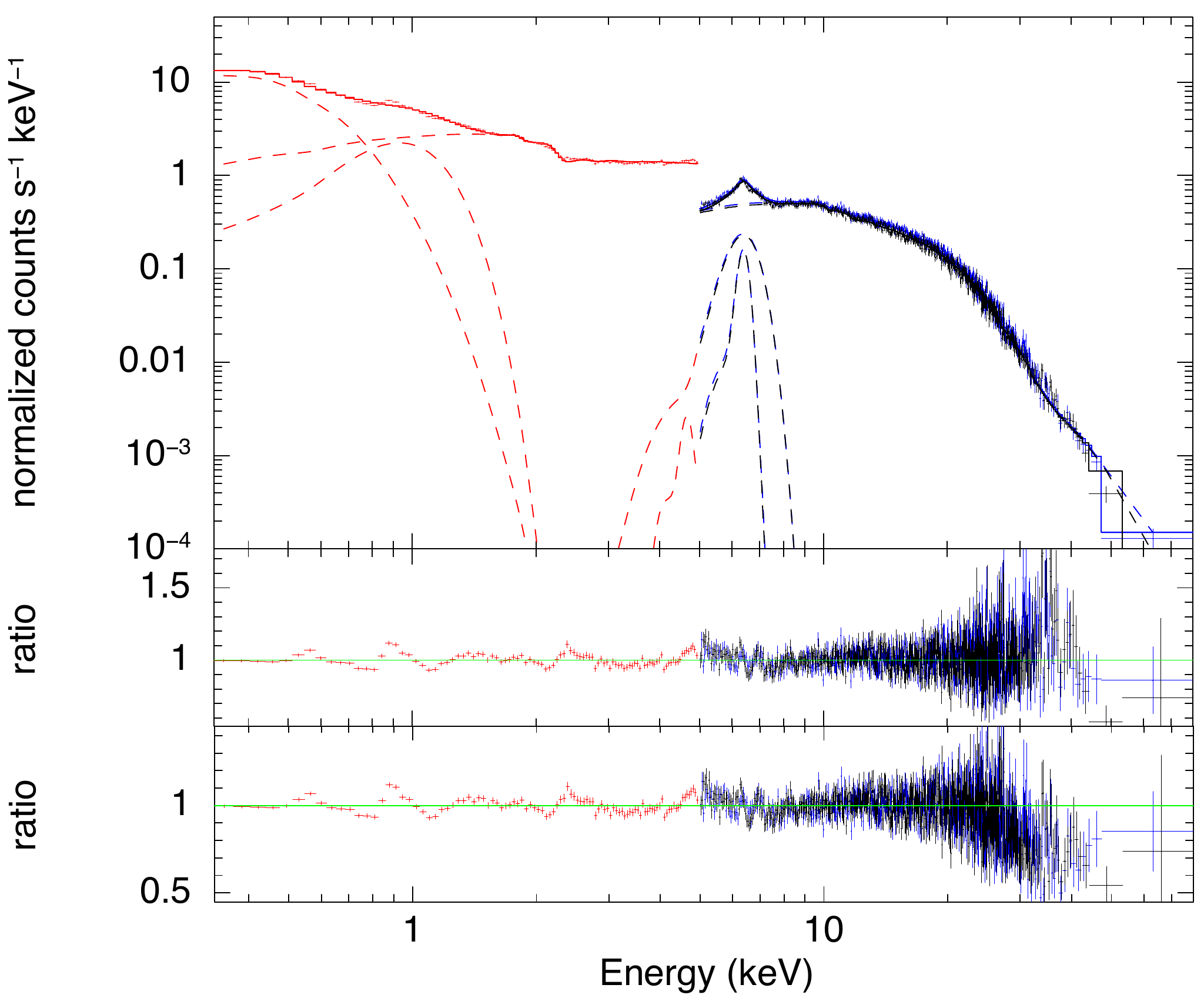} 
\includegraphics[scale=0.42]{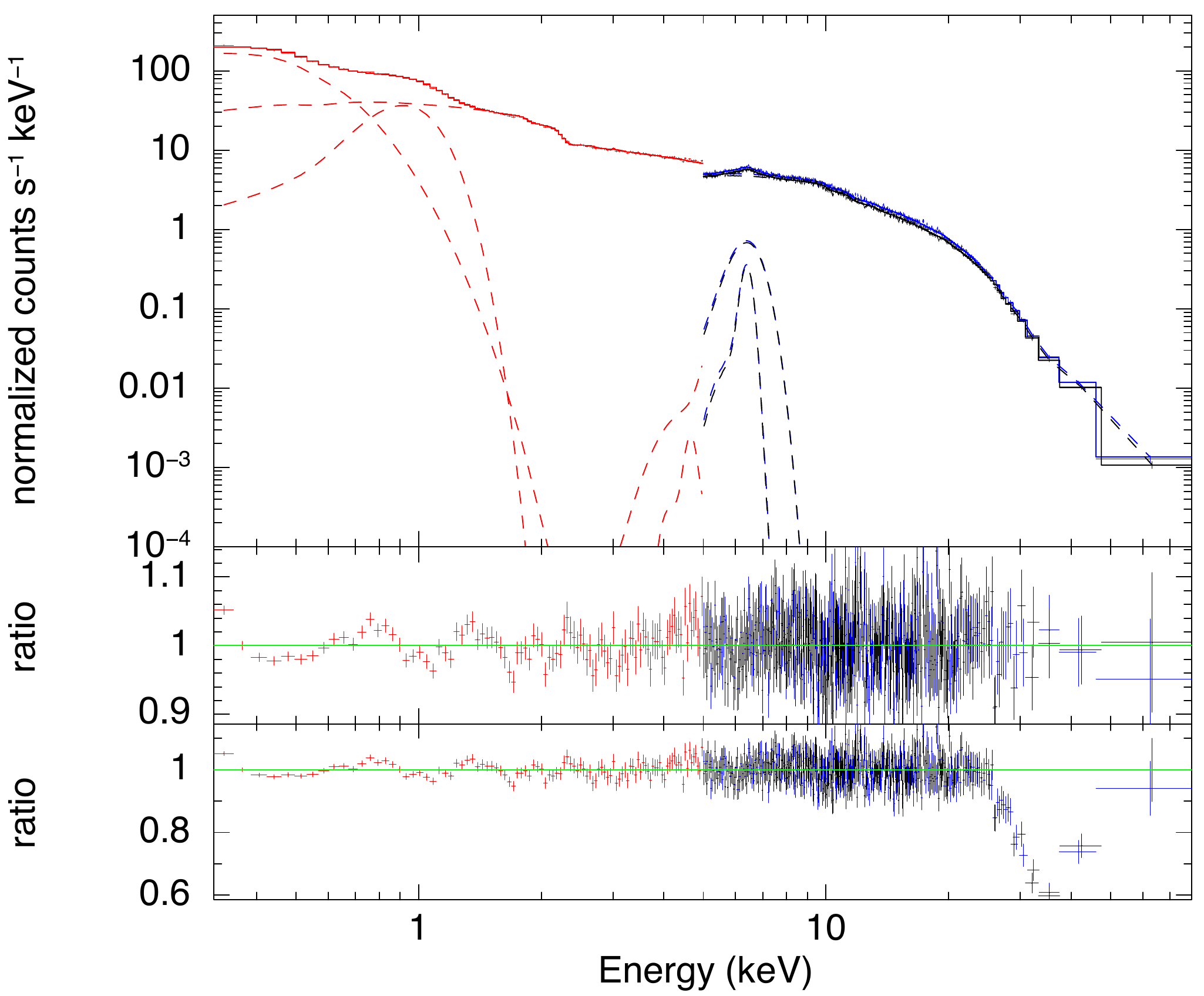}
\caption{Joint \xmm\ (red) and \nustar\ (FPMA - blue, FPMB - black) spectra for the Her X-1 observations H1 ($\phi_{\text{SO}}=0.22$, top left panel), H2 ($\phi_{\text{SO}}=0.33$, top right panel), H3 ($\phi_{\text{SO}}=0.74$, bottom left panel), and H4 ($\phi_{\text{SO}}=0.16$, bottom right panel). The \xmm\ spectrum is modeled from 0.3--5 keV while the \nustar\ spectra are modeled from 5--60 keV. For each spectrum, the top panel shows the spectrum and model components, the middle pannel shows the ratio of data to model for the best fit model, and the bottom panel shows the ratio of data to model with the CRSF feature removed.} Table \ref{tab:hnpexparams} contains the spectral parameters.
\label{fig:hspec}
\end{figure*}

We show the phase-averaged spectra and the residuals to our model fit for Observations H1, H2, H3, and H4 in Figure \ref{fig:hspec}. In Figure \ref{fig:hspec} we show the joint \xmm\ and \nustar\ spectrum in the first panel, the ratio of data to model for our best fit spectral model in the middle panel, and the ratio of data to model for our best fit model without the absorption feature representing the CRSF in the bottom panel. The bottom panels of Figure \ref{fig:hspec} demonstrate that the presence of a CRSF in the model is strongly preferred by the data for each observation. We note that Observations H2 and H3, which were fainter than the other two observations, have reduced signal at high energies which made the CRSF absorption model less well constrained compared to Observations H1 and H4. In order to fit our spectral model while including the CRSF, we fixed the values for line energy, width, and strength in Observations H2 and H3 to the average of the CRSF model parameters in Observations H1 and H4. Table \ref{tab:hnpexparams} contains the spectral parameters. 

We also calculated the flux of the entire modeled spectrum (0.3--60 keV) and the high energy band (8--60 keV) using the Xspec {\fontfamily{qcr}\selectfont flux} tool. We calculated the flux of the blackbody model component using Xspec model component {\fontfamily{qcr}\selectfont cflux}.

As in B20, we attempt to minimize differences in the \nustar\ and \xmm\ response functions by not modeling the spectra from these telescopes in their overlapping energy ranges. As we discussed in B20, this introduces somewhat artificial inflation to the calibration constants between the spectra. The differences in cross calibration are a known issue, albiet a poorly understood one, with the absolute calibration of \xmm\ in Timing Mode (e.g., B20). These values should not be taken as a reflection of the relative fluxes seen by \nustar\ and \xmm.

We do not present a spectral analysis of Observation R3 in this work, since it is presented with a comparable spectral model in \cite{ramsay2002}.

\begin{deluxetable*}{lcccc}
\label{tab:hnpexparams}
\tablecolumns{3}
\tablecaption{Her X-1 phase-averaged spectral parameters\tablenotemark{a}}
\tablewidth{0pt}
\tablehead{
\colhead{Parameter} & \colhead{Observation H1} & \colhead{Observation H2} & \colhead{Observation H3} &\colhead{Observation H4} }  
\startdata
Flux$_{\text{total}}$ (erg cm$^{-2}$ s$^{-1}$; 0.3--60 keV) & (6.418 $\pm$ 0.006)$\times 10^{-9}$   & (2.70 $\pm$ 0.02)$\times 10^{-10}$   & (9.72 $\pm$ 0.03)$\times 10^{-10}$ & (5.72 $\pm$ 0.01)$\times 10^{-9}$      \\
Flux$_{\text{power law}}$ (erg cm$^{-2}$ s$^{-1}$; 8--60 keV) & (4.559 $\pm$ 0.005)$\times 10^{-9}$ & (2.05 $\pm$ 0.01)$\times 10^{-10}$   & (7.67 $\pm$ 0.02)$\times 10^{-10}$ &(4.405 $\pm$ 0.01)$\times 10^{-9}$   \\
Flux$_{\text{blackbody}}$ (erg cm$^{-2}$ s$^{-1}$; 0.3--5 keV) &  (1.189 $\pm$ 0.003)$\times 10^{-10}$ & (2.54 $\pm$ 0.03)$\times 10^{-12}$  & (7.45 $\pm$ 0.05)$\times 10^{-12}$ &  (9.18 $\pm$ 0.06)$\times 10^{-11}$  \\
Photon Index 						&  0.933 $\pm$ 0.004				& 0.56 $\pm$ 0.01   & 0.26 $\pm$ 0.01    &0.933 $\pm$ 0.009	  \\
$A_{\text{powerlaw}}$               &(3.90 $\pm$ 0.02)$\times 10^{-2}$  &(1.30 $\pm$ 0.03)$\times 10^{-5}$ & (2.48 $\pm$ 0.03)$\times 10^{-5}$ & (3.34 $\pm$ 0.03)$\times 10^{-2}$  \\
Cut-off Energy (keV) 				&  19.8 $\pm$ 0.2 	                & 18.3 $\pm$ 0.4    & 17.8 $\pm$ 0.2     &19.4 $\pm$ 0.4  \\ 
Folding Energy (keV) 	            &  9.9 $\pm$ 0.2					& 9.0 $\pm$ 0.4     & 7.7 $\pm$ 0.2    &9.7 $\pm$ 0.3  	  \\
E$_{\text{CRSF}}$ (keV)				&  36.5 $\pm$ 0.4					& 36 (fixed)        & 36 (fixed)            &35.6 $\pm$ 0.7	\\
$\sigma_{\text{CRSF}}$ (keV)		&  5.1 $\pm$ 0.4					& 5 (fixed)         & 5 (fixed)            &4.8 $\pm$ 0.7		\\
CRSF Strength						&  6.6 $\pm$ 0.8				    & 6.5 (fixed)       & 6.5 (fixed)            &6 $\pm$ 2	\\
$kT_{\text{BB}}$ (keV) 				&  0.0919 $\pm$ 0.0003				& 0.090 $\pm$ 0.001 & 0.0932 $\pm$ 0.0008  &0.0884 $\pm$ 0.0006	  \\
$A_{\text{BB}}$ (keV) 				& (2.58 $\pm$ 0.01)$\times 10^{-3}$ & (5.7 $\pm$ 0.1)$\times 10^{-5}$ & (1.59 $\pm$ 0.02)$\times 10^{-4}$ &(2.09 $\pm$ 0.02)$\times 10^{-3}$ \\
E$_{\text{Fe K$\alpha$, broad}}$ (keV, fixed) 	& 6.4					& 6.4               & 6.4                 &6.4 	 \\
$\sigma_{\text{Fe K$\alpha$, broad}}$ (keV, fixed) & 0.5				& 0.5               & 0.5                 &0.5 	  \\
$A_{\text{Fe K$\alpha$, broad}}$ (photons cm$^{-2}$ s$^{-1}$)  	&(1.00 $\pm$ 0.09)$\times 10^{-3}$& (1.4 $\pm$ 0.3))$\times 10^{-4}$ & (1.03 $\pm$ 0.07)$\times 10^{-3}$ &(1.27 $\pm$ 0.2)$\times 10^{-3}$  \\
E$_{\text{Fe K$\alpha$, narrow}}$ (keV, fixed) 	    & 6.4				& 6.4               & 6.4                &6.4  \\
$\sigma_{\text{Fe K$\alpha$, narrow}}$ (keV, fixed) & 0.1				& 0.1               & 0.1                & 0.1 \\
$A_{\text{Fe K$\alpha$, narrow}}$ (photons cm$^{-2}$ s$^{-1}$)  	& (4.40 $\pm$ 0.04)$\times 10^{-4}$	& (5 $\pm$ 2)$\times 10^{-5}$ & (2.67 $\pm$ 0.03)$\times 10^{-4}$ &(2.3 $\pm$ 0.8)$\times 10^{-4}$   \\
$E_{\text{bump}}$ (keV, fixed) 		& 0.9							    & 0.9               & 0.9                &0.9 \\
$\sigma_{\text{bump}}$ (keV, fixed) & 0.191 $\pm$ 0.001			     	& 0.23 $\pm$ 0.01   & 0.243 $\pm$ 0.007  &0.173 $\pm$ 0.003 \\
$A_{\text{bump}}$ (photons cm$^{-2}$ s$^{-1}$) 	& (2.46 $\pm$ 0.02)$\times 10^{-2}$	& (5.6 $\pm$ 0.3)$\times 10^{-4}$ & (1.44 $\pm$ 0.05)$\times 10^{-3}$ &(1.58 $\pm$ 0.04)$\times 10^{-2}$\\*
$c_{\text{FPMA}}$ 					& 2.58 $\pm$ 0.02					& 1.14 $\pm$ 0.02   & 1.06 $\pm$ 0.02    &2.80 $\pm$ 0.04			 \\*
$c_{\text{FPMB}}$	 				& 2.64 $\pm$ 0.02                   & 1.16 $\pm$ 0.03   & 1.09 $\pm$ 0.02    &2.85 $\pm$ 0.04	 	 \\*
$c_{\text{EPIC-pn}}$ (fixed) 			& 1								& 1                 & 1                  &1	 \\*
\chisq\ 							& 2172.01							& 1179.42            & 1993.86            &872.82	  \\*
Degrees of Freedom 				    & 1260							    & 1035               & 1345               &675	
\enddata
\tablenotetext{a}{For the continuum model {\fontfamily{qcr}\selectfont constant * tbnew * (powerlaw * highecut * gabs * gabs + bbody + gauss + gauss + gauss)}. The errors on the flux are 1$\sigma$ and the errors on the parameters are 90\% confidence intervals.}
\tablenotetext{b} {We select 8--60 keV for the power law flux because this energy range is consistent with the hard band used in our timing analysis. }
\end{deluxetable*}

\section{Results} \label{Hdisksec:results}

\subsection{Pulse Profiles}

Her X-1's pulse profiles from Observations H1 and H4 (Figure \ref{fig:herpp}) are somewhat similar in shape and relative phase. The hard \nustar\ profile has a strong, narrow main peak and a weaker, broader second peak. The soft \xmm\ pulses have a broad, single-peaked structure. The main peaks of the hard and soft profiles are almost 180\degree\ out of phase in both Observation H1 and Observation H4. 

The hard pulse profile from Observation R3 has a similar shape to the hard pulses in observations H1 and H4. However, the shape of the soft pulse changes, with the pulse steadily building strength up to the pulse maximum and then dropping off rapidly, as opposed to the rapid rise and slow decline of the pulses in Observations H1 and H4. There is also a change in the relative phase between the maxima of the hard and soft pulsations. The peaks of the hard and soft pulse profiles are separated by approximately 0.5 phase in Observations H1 and H4 and by approximately 0.2 phase in Observation R3.

The differences between these three pulse profiles demonstrate that the shape and relative phase of the soft pulses changes with superortibal cycle. The similarity in the relative phase of the pulse profiles from Observations H1 and H4, which both take place around $\phi_{\text{SO}} \approx 0.2$, shows that the pulses return to their original configuration after a complete precession cycle. However, if we create pulse profiles with finer binning we can also see differences in the fine structure of the hard pulse profiles that arise from the difference in superorbital cycle between Observation H1 and H4 (e.g., \citealt{fuerst2013,staubert2013}). In Figure \ref{fig:highrespp} we show the hard band pulse profile from Observations H1 and H4 made with 128 bins per phase. Both pulse profiles show a notch in the bright main pulse as well as structure within the interpulse. There are small differences in the fine structure of these profiles, which become even more noticeable in the energy-resolved pulse profiles in Appendix A, but these differences are consistent with expectations from the slightly different superorbital phase of these observations, based on the \cite{staubert2013} template. Notably, Observation H4's interpulse contains two distinct bump like features, where typically only one is observed, however such fluctuations have been occasionally seen in Her X-1's pulse profiles with no obvious correlation to superorbital phase (\citealt{staubert2013}) . By showing simulated pulse profiles (see Section \ref{Hdisksec:diskmodel}) in Figure \ref{fig:highrespp} with similar phase binning we demonstrate that the warped disk model is not capable of reproducing the fine structure in the pulse profiles. For this reason, we use coarse binned pulse profiles throughout our warped disk modeling procedure. 

\subsection{Spectroscopy}

Our spectroscopic analysis indicates that an absorbed power law and soft blackbody component is a good description of the broadband X-ray continuum of Her X-1. Because Observations H1 and H4 capture a similar phase in the precession cycle of the inner disk, we expect that the spectra from these observations should also be similar. We do find several similarities in the spectral parameters (Table \ref{tab:hnpexparams}); the power law parameters, including photon index, cut-off energy, and folding energy are consistent between the two observations. We also note that our fits for these two observations are consistent with those presented in \cite{staubert2020}. There are some small differences between the blackbody temperature and strength or normalization of the emission line components between these two observations. These differences may be expected given the small difference in flux between these observations, which can be seen in the difference between pulse profile count rates in Observations H1 and H4 (see Fig. \ref{fig:herpp}).

Our spectral fits to Observations H2 and H3 show some distinct differences in the power law shape from Observations H1 and H4. Although some difference in spectral shape with superorbital phase is expected (\citealt{fuerst2013}), Observations H2 and H3 have signficiantly lower flux than Observations H1 and H4. This is consistent with obscuration of the neutron star by the accretion disk, which would also explain other features of these observations, including the differences in spectral continuum shape and pulse strength (\citealt{deeter1998,leahy2000}). Increased obscuration could also explain the weakened CRSF present in Observation H2 and H3, since the reduced flux of the entire spectrum would make this feature more difficult to detect. The non-detection of the CRSF at low flux is supported by previous work by \cite{inam2005}, who did not find a CRSF feature necessary to fit the low state spectrum of Her X-1.

\begin{figure*}
\centering
\includegraphics[scale=0.6]{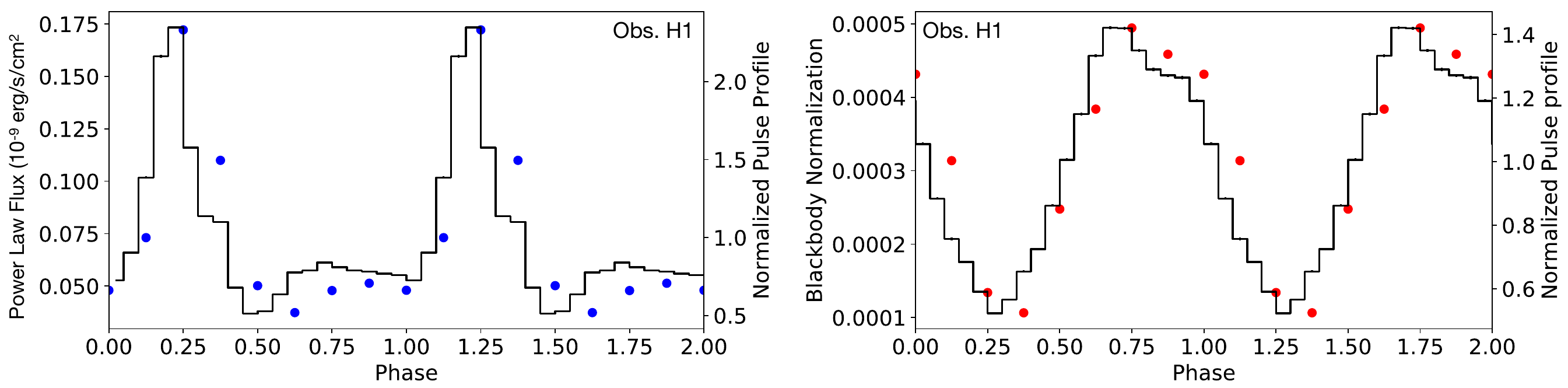} \\
\includegraphics[scale=0.6]{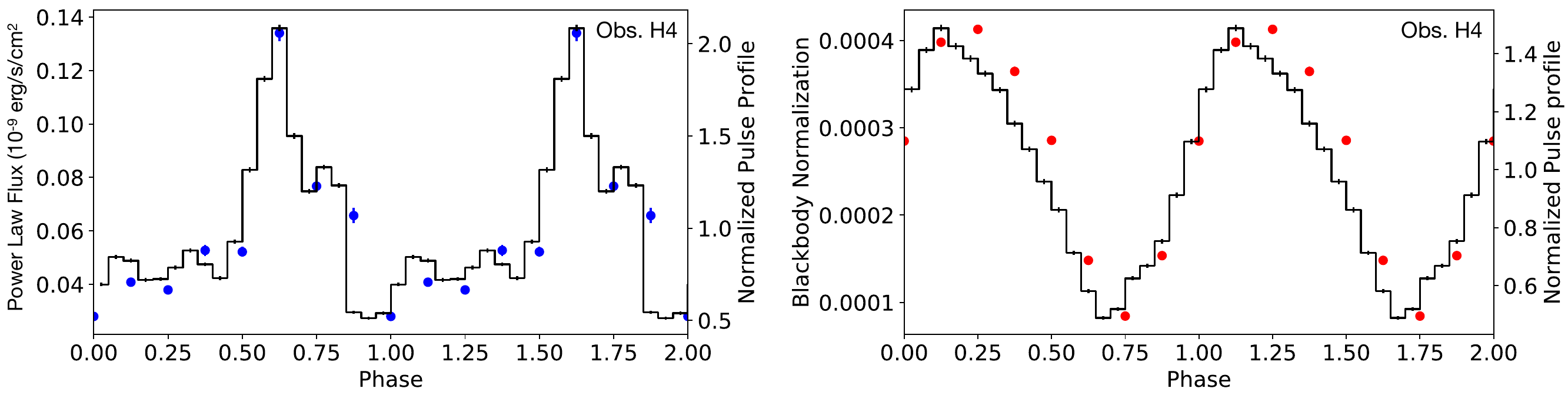}
\caption{Left: Power law flux (8--60 keV, blue points) compared to the \nustar\ 8--60 pulse profile for Observation H1 (top row, $\phi_{\text{SO}}=0.20$) and Observation H4 (bottom row, $\phi_{\text{SO}}=0.14$). Right: Blackbody normalization (red points) compared to the \xmm\ 0.3--0.7 keV pulse profile for Observation H1 (top row) and H4 (bottom row). In both cases, the phase-resolved spectroscopy parameters are in good agreement with their respective pulse profiles. The agreement between the spectral parameters and their respective energy resolved pulse profiles demonstrates that the pulse profiles are a suitable proxy for their respective flux parameters in our warped disk model.}
\label{fig:hfluxpp}
\end{figure*}

\section{Modeling the warped inner disk} \label{Hdisksec:diskmodel}

To constrain the geometry of the inner accretion disk during its precession we used the same model as B20, which was first presented in \cite{hickoxvrtilek2005}. A full description of the model, its underlying assumptions, and a schematic diagram can be found in B20. We briefly describe an important assumption here for clarity.

A fundamental assumption of the B20 warped disk model is that the emission below 1 keV follows the blackbody emission of the accretion disk and that hard X-ray emission follows the power law. We demonstrate this relationship by performing a simplistic phase resolved spectral analysis. For Observations H1 and H4, where strong pulsations were detected, we used the HENDRICS tool {\fontfamily{qcr}\selectfont HENphasetag} to assign pulse phase values to each photon in the \nustar\ and \xmm\ event files. We then filtered the data into 8 equal phase bins using {\fontfamily{qcr}\selectfont xselect} for \nustar\ data and {\fontfamily{qcr}\selectfont evselect} for \xmm\ data. We extracted spectra and grouped them with a minimum of 100 counts per bin. We modeled the joint phase-resolved spectra between 0.3--47 keV due to the poorer statistics in these spectra.

We attempted to use the phase-averaged spectral model to also describe the phase-resolved spectra, but the reduced signal to noise in the phase-resolved spectra required that we reduce degeneracy between some model parameters. We fixed the blackbody temperature, the width of the CRSF, and the width of the 0.9 keV bump feature to their phase averaged values. We also removed the broad iron line model component and only used the narrow emission line at 6.4 keV. This simplified spectral model allowed us to evaluate the changes in the blackbody and power law normalizations.

In Figure \ref{fig:hfluxpp} we show the power law flux and the \nustar\ 8--60 keV pulse profile as well as the blackbody normalization compared to the \xmm\ 0.3--0.7 keV pulse profile from Observation H1. Both the hard and the soft pulses show good phase agreement. The good agreement between the integrated flux in this energy range and their respective pulse profiles demonstrates that we can use the pulse profiles as a suitable proxy for the spin-resolved power law and blackbody flux. Using this proxy, we can use the warped disk model to simulate the energy resolved \xmm\ and \nustar\ pulse profiles and assume that the profiles follow any changes in the respective strength of the blackbody and power law.

With this relationship verified, we will now describe the warped disk model itself. This model uses a simple warped disk geometry, represented by a series of concentric circles that are tilted and twisted relative to one another, to describe the inner accretion disk. The disk is defined by the radius and tilt angle of the inner ($r_{\text{in}}$,$\theta_{\text{in}}$) and outer ($r_{\text{out}}$,$\theta_{\text{out}}$) rings and their relative twist angle with respect to one another ($\theta_{\text{tw}}$). The height of the observer is set by the observer angle ($\theta_{\text{obs}}$).

The pulsar emission geometry can be represented either as a narrow pencil beam or by a wider fan beam. The beam geometry consists of two beams, whose location on the neutron star surface are defined by $\theta_{\text{b}}$ and $\phi_{\text{b}}$, which are the angle out of the rotational plane and the azimuthal angle, respectively. The pencil beam profile is a two-dimensional Gaussian with width $\sigma_{\text{b}}$. The fan beam is also a two dimensional Gaussian with an additional opening angle ($\theta_{\text{fan}}$). 

Once the disk and emission geometry are specified, the model calculates the simulated pulse profile at 30 pulse phases and 8 disk precession angles. As the pulsar rotates, the beam profile irradiates the inner accretion disk. The hard pulse profile is made by calculating the luminosity of the beam visible to the observer as a function of pulse phase and disk precession angle. We generate the soft pulse profile by calculating the luminosity of the irradiated disk visible to the observer. We then compare the simulated pulse profiles to the observed profiles. This model does not include the effects of general relativity or light bending.

When modeling Her X-1's inner disk, we are able to use previous models of Her X-1's warped disk to guide our choice of parameters. \cite{scott2000} created a model disk that would reproduce observed changes in Her X-1's observed pulse profiles between the main on and short on of the superorbital cycle. They used a low observer elevation consistent with the high inclination of the source ($\theta_{\text{obs}}=-5$\degree) and defined the disk as having an inner tilt angle ($\theta_{\text{in}}$) of 11\degree and an outer tilt angle ($\theta_{\text{out}}$) of 20\degree. Furthermore, they used a twist angle of 139\degree between the inner and outer rings of the disk. Similar parameters were used by \cite{leahy2002}, who created a warped disk to reproduce the observed 35 day supeorbital modulation as seen by \textit{RXTE}. \cite{leahy2002} found an the observer angle of 5\degree\ and an outer tilt angle to 30\degree\ best reproduced the shape of the superorbital cycle. For this work, we adopted the disk parameters found by \cite{scott2000} and used an observer elevation of -5\degree. We also tried using the \cite{scott2000} outer disk tilt angle of 20\degree\ but ultimately found a better fit to our observed pulse profiles with the \cite{leahy2002} value of 30\degree. Our full set of model parameters are shown in Table \ref{tab:hdiskpar}.

\begin{figure*}
\centering
\begin{tabular}{c}
\includegraphics[scale=0.8]{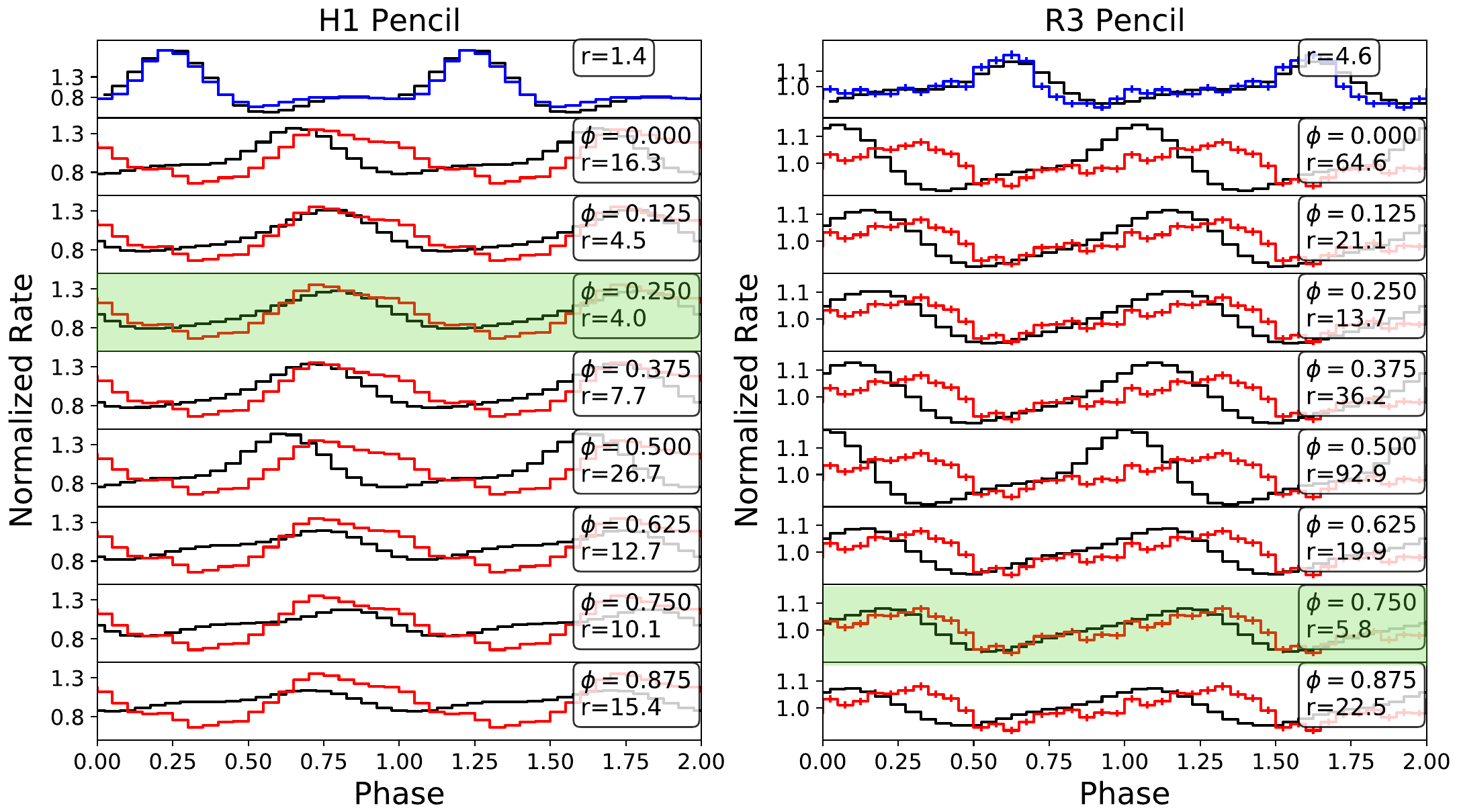} \\
\includegraphics[scale=0.8]{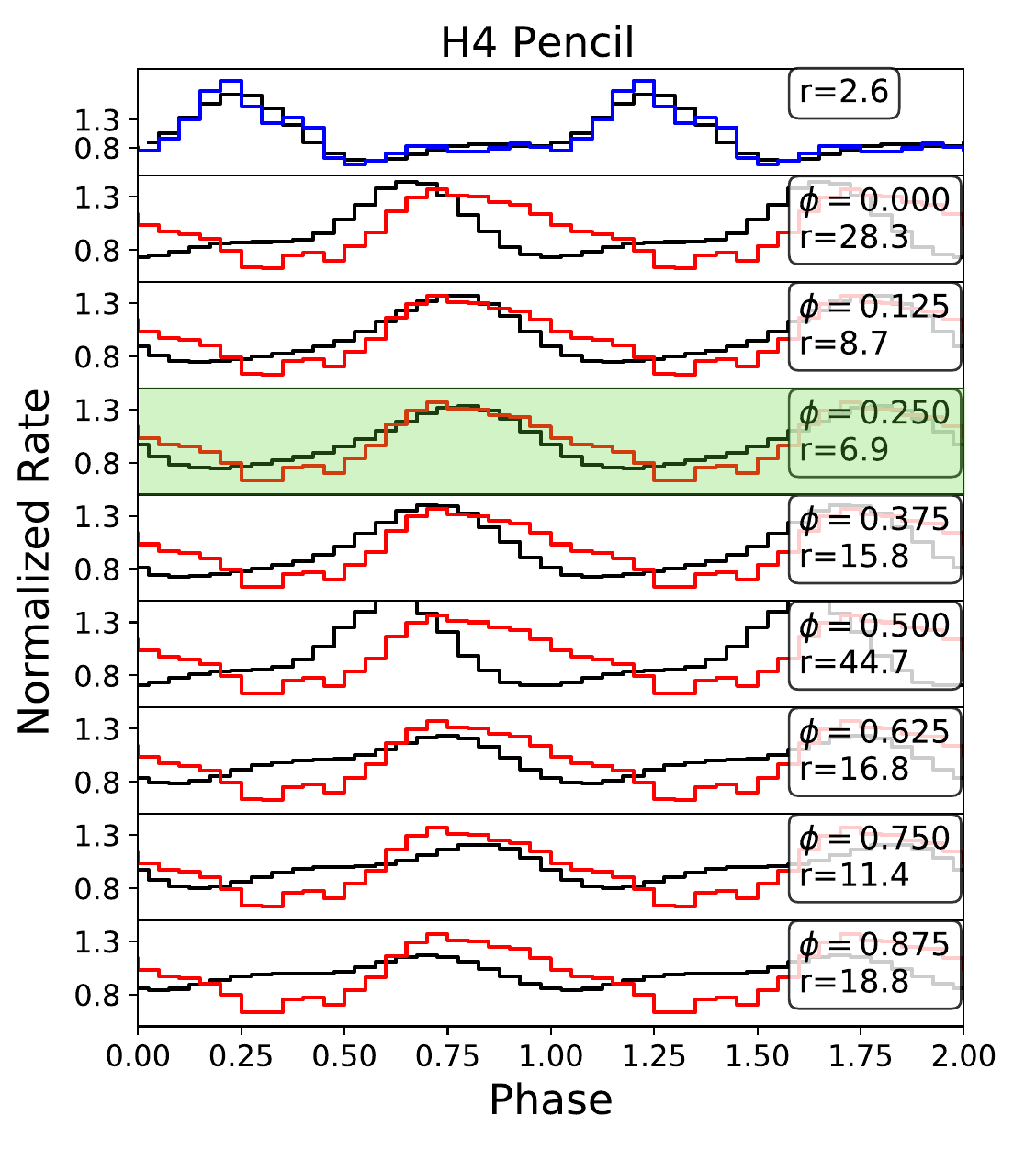}
\end{tabular}
\caption{Observed hard (blue) and soft (red) pulse profiles compared with simulated (black) pulse profiles from the warped disk model with a pencil beam for the three Her X-1 observations. The disk precession angles ($\phi$) correspond to the 35 day superorbital phase. For each disk precession angle, we calculate the goodness of fit with the parameter r ($r=\sum(P_{\text{obs}}(\phi_{\text{spin}}) - P_{\text{sim}}(\phi_{\text{spin}})) / \overline{P_{\text{obs}}}$, where $P_{\text{obs}}$ is the observed pulse profile and $P_{\text{sim}}$ is the simulated pulse profile)}. For the soft pulses, a disk precession of $\phi_{\text{SO}}=0.25$ (highlighted in green) best describes the observed pulse profiles of Observations H1 and H4, while a disk precession of $\phi_{\text{SO}}=0.75$ best describes Observation R3.
\label{fig:hdisksimpppencil}
\end{figure*}

\begin{figure*}
\centering
\begin{tabular}{c}
\includegraphics[scale=0.8]{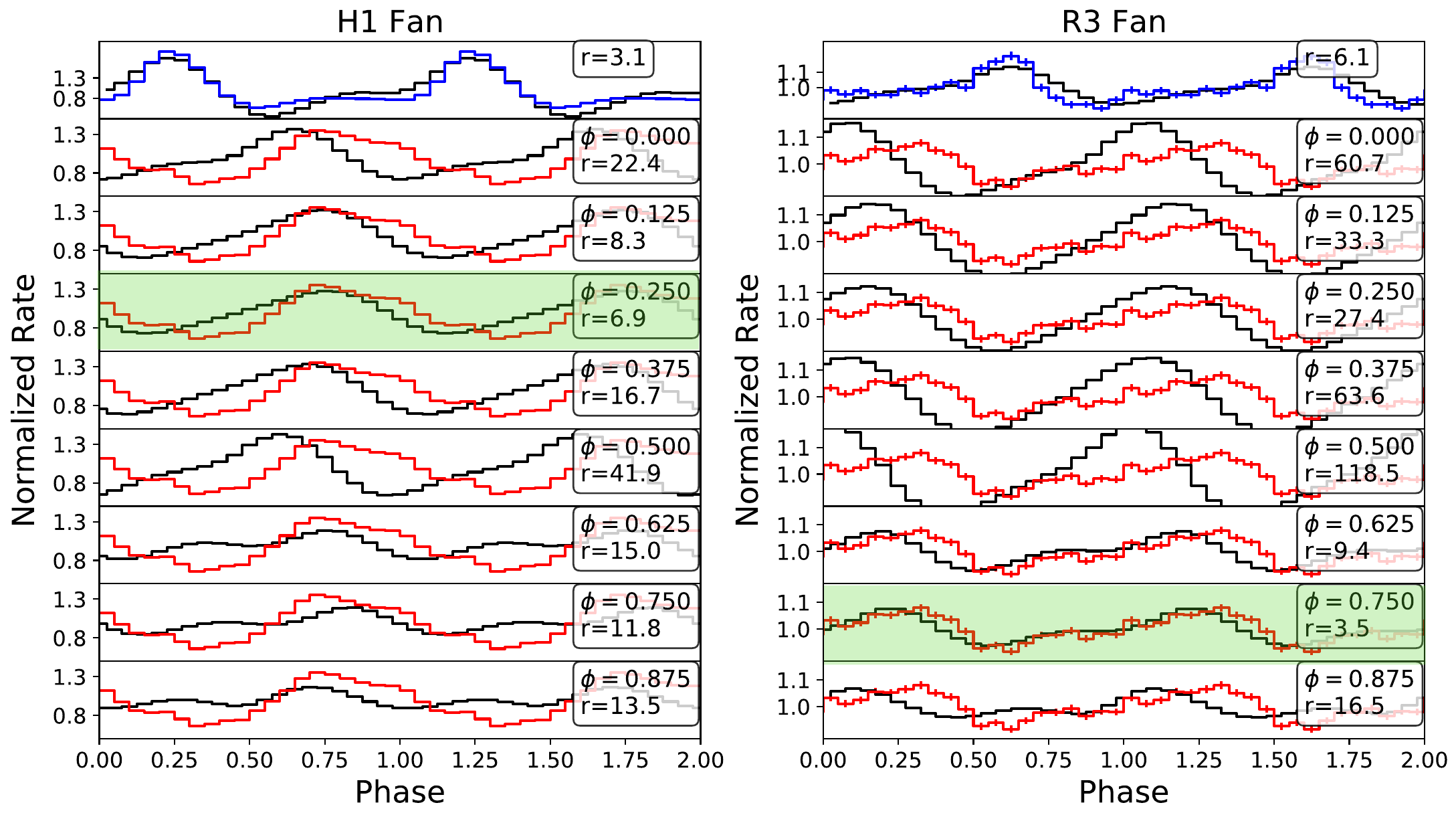} \\
\includegraphics[scale=0.8]{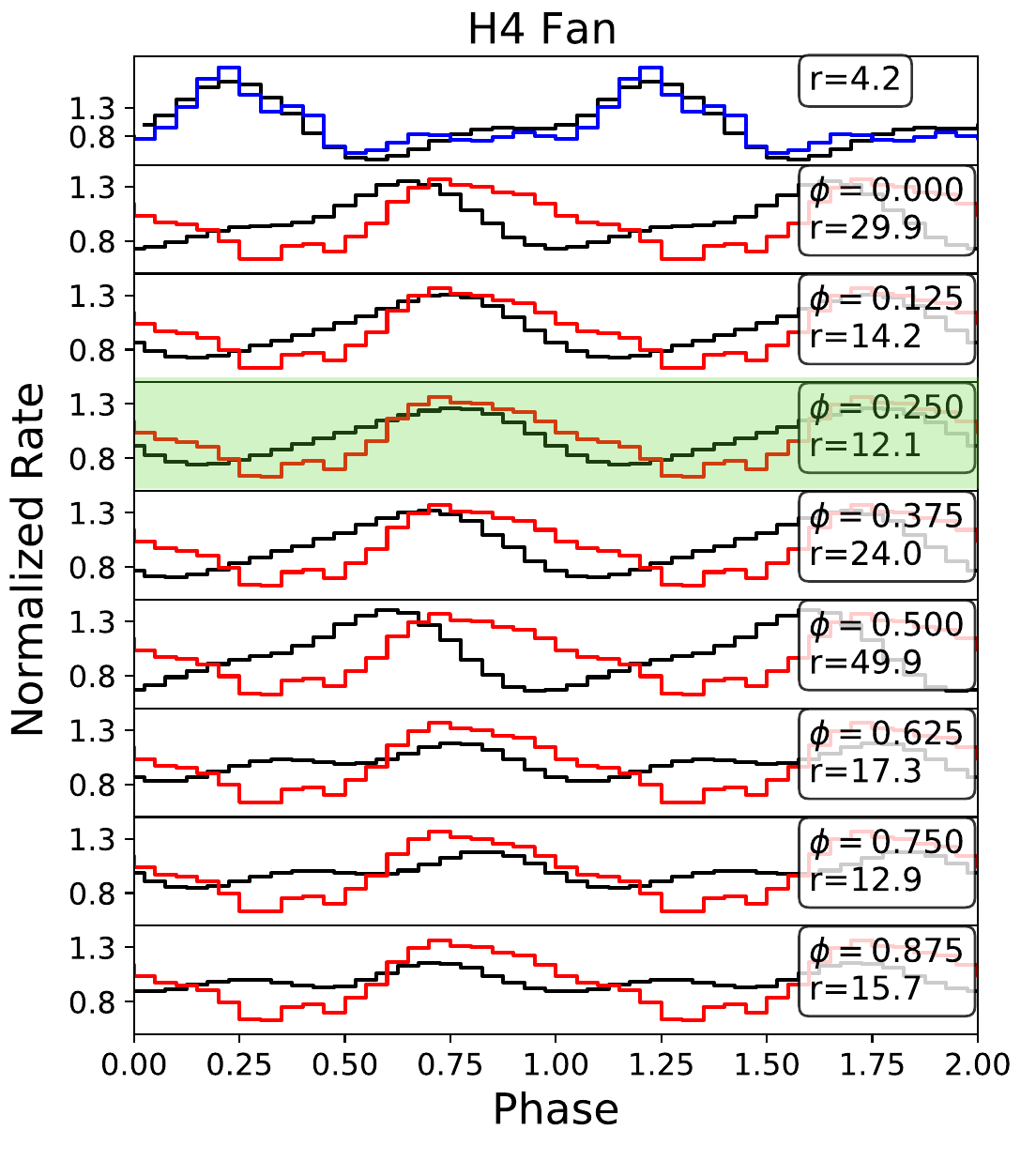}
\end{tabular}
\caption{Observed hard (blue) and soft (red) pulse profiles compared with simulated (black) pulse profiles from the warped disk model with a fan beam for the three Her X-1 observations. The disk precession angles ($\phi$) correspond to the 35 day superorbital phase. For each disk precession angle, we calculate the goodness of fit with the parameter r ($r=\sum(P_{\text{obs}}(\phi_{\text{spin}}) - P_{\text{sim}}(\phi_{\text{spin}})) / \overline{P_{\text{obs}}}$, where $P_{\text{obs}}$ is the observed pulse profile and $P_{\text{sim}}$ is the simulated pulse profile)}. For the soft pulses, a disk precession of $\phi_{\text{SO}}=0.25$ (highlighted in green) best describes the observed pulse profiles of Observations H1 and H4, while a disk precession of $\phi_{\text{SO}}=0.75$ best describes Observation R3.
\label{fig:hdisksimppfan}
\end{figure*}

We kept the disk geometry set to the values described above for all three observations. We allowed the location of the two beams and their widths to vary until we matched the observed shape of the hard pulse profiles (Table \ref{tab:hdiskpar}). We allowed the model disk to precesss around the pulsar, which changed the shape and phase of the soft pulse profile component. We calculated the simulated hard and soft pulse profiles at eight equally spaced disk precession angles. We then fit the simulated pulse profile to the observed pulse profile, allowing the overall amplitude of the simulated pulsations to scale, and estimated the goodness of fit by calculating $r=\sum(P_{\text{obs}}(\phi_{\text{spin}}) - P_{\text{sim}}(\phi_{\text{spin}})) / \overline{P_{\text{obs}}}$, where $P_{\text{obs}}$ is the observed pulse profile and $P_{\text{sim}}$ is the simulated pulse profile. The disk precession phase with the lowest $r$ value represents the orientation of the disk that reproduces the observed pulsations best, and these phases are highlighted in green in Figures \ref{fig:hdisksimpppencil} and \ref{fig:hdisksimppfan}. We performed the fits for the pencil and fan beam configurations independently.

\begin{figure*}
\centering
\includegraphics[scale=0.7]{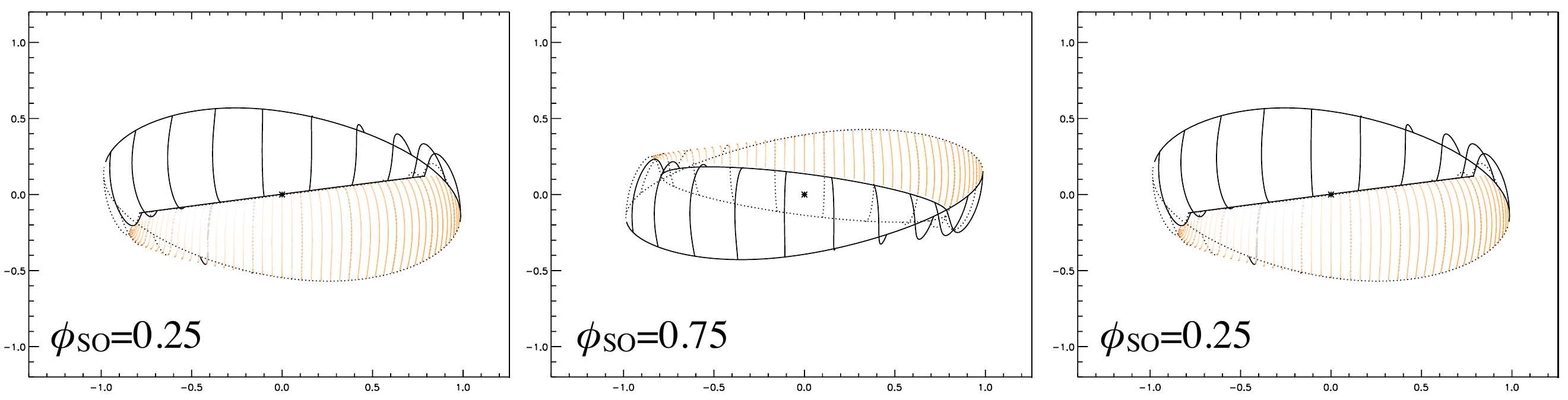} 
\caption{The simulated disk that best reproduces the observed pulse profiles from Observation H1, R3, and H4 for both the fan beam and pencil beam models. The orange shaded section of the disk represents the illuminated side of the disk, while black lines indicate the back of the disk which is not illuminanted by the pulsar beam. Units are 10$^{8}$ cm.}
\label{fig:hdiskim}
\end{figure*}

Because the energy-resolved pulse profiles for Observation H1 and Observation H4 are extremely similar, the best fit beam geometry and disk precession angle is the same in both observations. The disk precession angle of 0.25 closely matches the superorbital phase of Observations H1 and H4 ($\phi_{\text{SO}} \approx 0.2$). In Observation R3, the disk precession angle that best reproduced the observed pulse profiles was slightly off, at angle 0.75 rather than the closer value of 0.625. We are encouraged by the fact that the model came close to selecting the correct phase bin, especially considering that Observation R3 occurred almost 18 years before Observations H1 and H4. We show the model disks that produced our best fit simulated pulse profiles in Figure \ref{fig:hdiskim} and list the parameters for the disk and beam geometries in Table \ref{tab:hdiskpar}.

\begin{deluxetable*} {ccccccc} 
\label{tab:hdiskpar}
\tablecolumns{5}
\tablecaption{Disk Model Parameters}  
\tablewidth{0pt}
\tablehead{ \multicolumn{1}{c}{}  & \multicolumn{2}{c}{Observation H1}  & \multicolumn{2}{c}{Observation R3} & \multicolumn{2}{c}{Observation H4} \\ 
\cmidrule(lr){2-3} \cmidrule(lr){4-5} \cmidrule(lr){6-7}\\ 
\colhead{Parameter} & \colhead{Pencil Beam} & \colhead{Fan Beam} & \colhead{Pencil Beam} & \colhead{Fan Beam} & \colhead{Pencil Beam} & \colhead{Fan Beam} }
\startdata
$r_{\text{in}}$ (10$^{8}$ cm) & 0.8 & 0.8 & 0.8 & 0.8 & 0.8 & 0.8 \\
$r_{\text{out}}$ (10$^{8}$ cm) & 1 & 1 & 1 & 1 & 1 & 1 \\
Inner tilt $\theta_{\text{in}}$ (\degree) & 10 & 10 & 10 & 10 & 10 & 10\\
Outer tilt $\theta_{\text{out}}$ (\degree) & 30 & 30 & 30 & 30 & 30 & 30\\
Twist angle $\phi_{\text{tw}}$ (\degree) & 139 & 139 & 139 & 139 & 139 & 139\\
Beam$_{1}$ angle from rotational plane $\theta_{\text{b}1}$ (\degree) 	& 0	& 40  & 0 &	40 & 0   & 40 \\
Beam$_{2}$ angle from rotational plane $\theta_{\text{b}2}$ (\degree) 	 & 60	&  60 & 60 & 60	& 60   & 60 \\
Beam$_{1}$ azimuth $\phi_{\text{b}1}$ (\degree) & 0 & 0 & 0 & 0 & 0 & 0 \\
Beam$_{2}$ azimuth $\phi_{\text{b}2}$ (\degree) & 	210 &  140	& 220 & 130 & 210  &140  \\
Beam half-width $\sigma_{\text{b}}$ (\degree) & 45	&  60	& 45, 60\tablenotemark{a} & 60 & 45   &60  \\
Fan beam opening angle $\theta_{\text{fan}}$ (\degree) & 0 & 60 & 0 & 60 & 0 &  60 \\
Observer elevation $\theta_{\text{obs}}$ (\degree) & -5 & -5 & -5 & -5 & -5 & -5
\enddata
\tablenotetext{a}{Here the beam width were asymmetric, with $\sigma_{\text{b}1}$=45\degree\ and $\sigma_{\text{b}2}$=60\degree.}
\end{deluxetable*}

\section{The pre-eclipse dip} \label{Hdisksec:dip}
Observation H4 captured a pre-eclipse dip with both \nustar\ and \xmm, as seen in the light curve in Figure \ref{fig:herlc}. This observation began at 58556.278576 MJD and ended at 58556.750799 MJD, with the next eclipse occurring at 58556.874225 MJD. \cite{giacconi1973} noted the presence of pre-eclipse dips in long term UHURU light curves or Her X-1. During these dips the flux of Her X-1 drops significantly as most of the pulsating emission becomes heavily absorbed (\citealt{giacconi1973,stelzer1999}). The unabsorbed emission spectrum is well described by a power law, suggesting that this persistent emission results from X-ray scattering in the obscuring matterial (\citealt{vrtilek1985,choi1994,leahy1994,reynolds1995}). Modeling of long term X-ray light curves suggest that these dips are caused by obscuration of material at the impact region of the accretion stream and accretion disk (e.g., \citealt{igna2012}).

\begin{figure*}
\centering
\includegraphics[scale=1.0]{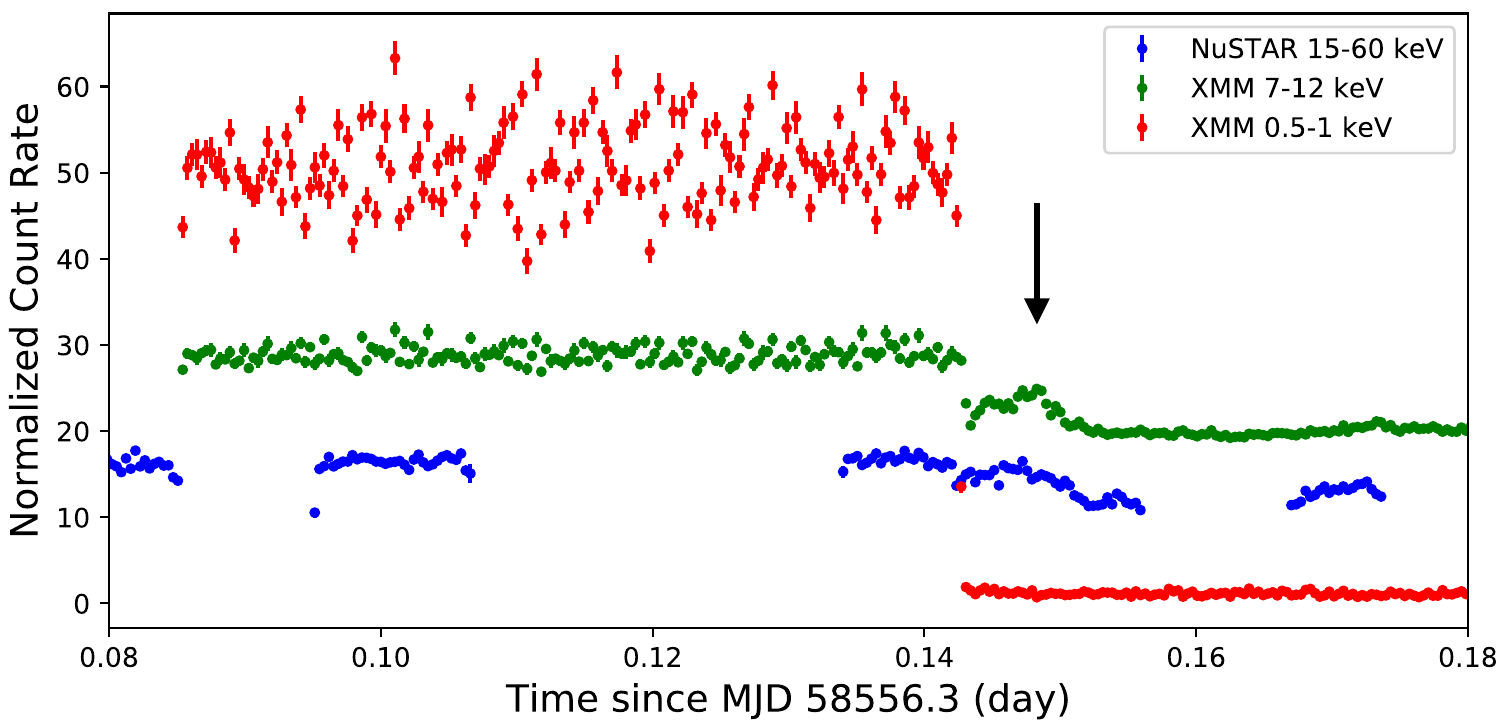} 
\caption{Energy resolved light curves centered around the onset of the pre-eclipse dip in Observation H4 and binned with 30 second bins. The transition is abrupt in the soft X-rays (0.5--1 keV, red) and more gradual in the mid (7--12 keV, green) and hard (15--60 keV, blue) curves. The black arrow marks a re-brightening event during the onset of the dip that is visible in the higher energy bands. The count rates for each light curve have been arbitrarily offset for clarity. We cleaned the \xmm\ light curves using the SAS tool {\fontfamily{qcr}\selectfont epiclccorr} to remove bins with low exposure fraction due to Counting Mode. }
\label{fig:hdiplc}
\end{figure*}

This dip presented an opportunity to examine the onset of the dip as a function of energy across a wide range of X-ray energies. In Figure \ref{fig:hdiplc} we show energy-resolved light curves of the onset of the dip, binned by 30 seconds. In the soft (0.5-1 keV) X-rays, the transition from bright to faint emission is almost immediate, taking place in a single 30 s time bin. In the middle (7-12 keV) and hard (15-60 keV) X-rays the transition occurs more slowly, over hundreds of seconds. The varying response of the energy resolved light curves is consistent with the generally held picture of an increase in absorber column density and scattering of the hard X-ray continuum (e.g., \citealt{vrtilek1985,leahy1994}).

The black arrow in Figure \ref{fig:hdiplc} marks a short re-brightening event seen during the onset of the pre-eclipse dip in the higher energy X-ray bands. This event could be related to the ``spike" phenomenon first seen by \cite{vrtilek1985}, where short increases in X-ray luminosity were seen during Her X-1's pre-eclipse dips. \cite{vrtilek1985} reported that these spikes reached as much as 80\% of the  pre-dip flux, lasted about 5--8 minutes, and repeated with a period of 108 minutes.

\begin{figure*}
\centering
\includegraphics[scale=0.7]{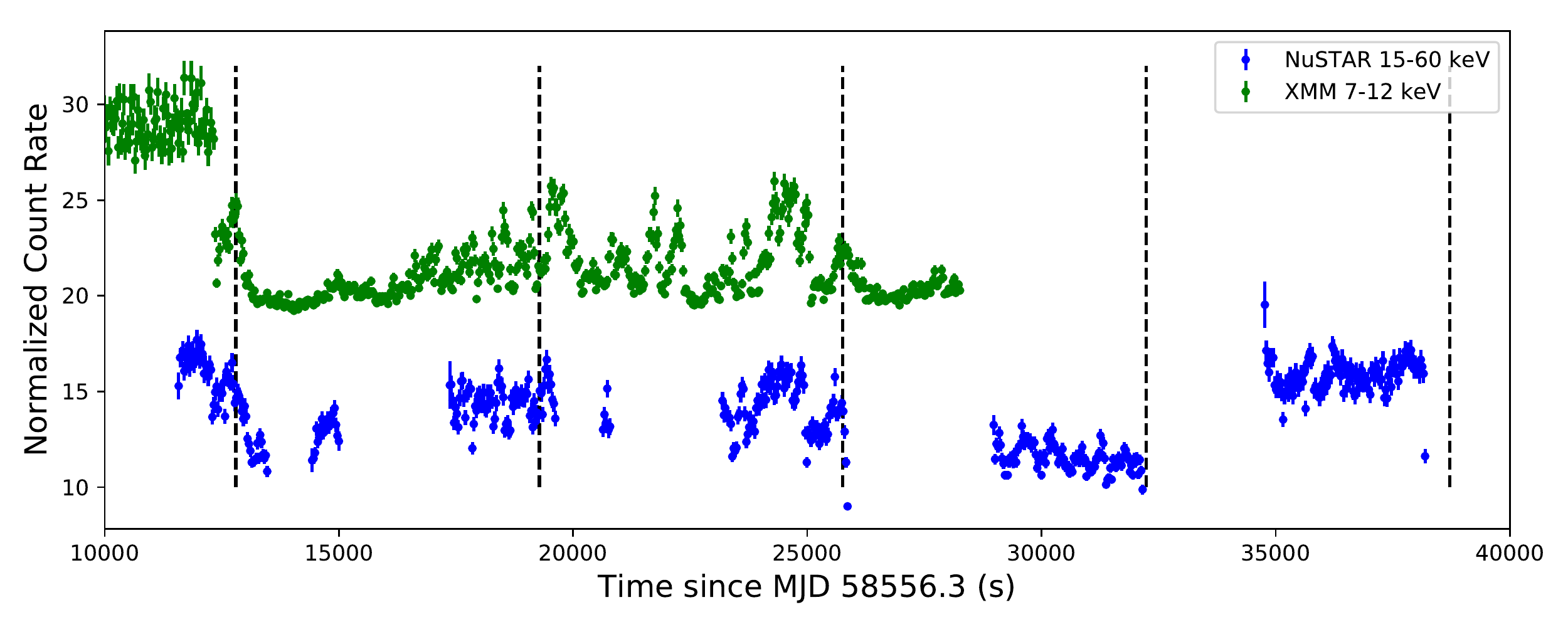} 
\caption{The same middle (7--12 keV, green) and hard (15--60 keV, blue) light curves from Figure \ref{fig:hdiplc}, but shifted in time to focus on the dip, rather than its onset. The onset of the dip starts at 12000 s in this plot. The black dashed lines are placed at 108 minute intervals, starting at the first re-brightening event during the onset of the dip. The activity seen in this pre-eclipse dip does not appear to follow a 108 minute period.}
\label{fig:hspikes}
\end{figure*}

While the soft (0.5--1 keV) energy band is mostly obscured during the pre-eclipse dip, we do see variations in brightness within the dip in the middle (7-12 keV) and hard (15-60 keV) energy bands, which we show in Figure \ref{fig:hspikes}. We performed an epoch folding search of the dip light curve to check for periodic behavior and found no significant periods within the pre-eclipse dip.

To further check for the 108 minute period found by \cite{vrtilek1985}, we included vertical dashed lines in Figure \ref{fig:hspikes} at 108 minute intervals starting with the re-brightening event marked with the black arrow in Figure \ref{fig:hdiplc}. While the first, second, and third interval do align with some variations in the light curve, there is significantly more variation present than can be described with a 108 minute period. Additionally, the variations we see in this dip appear less pronounced than those seen by \cite{vrtilek1985}.

While we do not see the clear spiking phenomenon observed by \cite{vrtilek1985}, we do see significant variability in the hard X-ray flux during the pre-eclipse dip. This variability seems consistent with that observed by \cite{leahy1994} and could possibly be caused by irregularities in the structure of the obscuring material. We leave a more detailed analysis of the variations seen in Observation H4 and their spectral similarity to the spikes of \cite{vrtilek1985} to a later analysis.

\section{Discussion} \label{Hdisksec:disc}

Several previous works including \cite{mccray1982}, \cite{scott2000}, \cite{leahy2002}, \cite{ramsay2002}, \cite{zane2004}, \cite{hickox2004}, \cite{kuster2005}, and \cite{staubert2013} suggested that the changes in pulse profile shape with superorbital phase in Her X-1 were caused by reprocessing in the inner accretion disk during its precession around the neutron star. In this work we use three observations of Her X-1 at different superorbital phases to show that the observed changes in pulse profile shape and relative phase can be modeled by a simple precessing accretion disk. 

If disk precession is the cause of the changes in shape and phase of the pulse profiles, then we expect that observations from the same superorbital phase should have similar pulse profile shapes. We confirm this expectation with the Observation H1 and H4 pulse profiles, which have similar pulse shapes and the same relative phase offset between the hard and soft pulsations. These results are strengthened by good agreement with archival data of Her X-1, particularly the \nustar\ pulse profiles presented in \cite{fuerst2013} from superorbital phases 0.11 and 0.17 which show a similar pulse shape to H1 and H4. 

We also expect to see periodicity in the spectral continuum with superorbital cycle in the precessing disk scenario. In our joint \xmm\ and \nustar\ spectral analysis we do see similarities between Observations H1 and H4, particularly in the high energy continuum (photon index, power law cutoff and folding energies), the shape of the CRSF, and the size of the soft bump feature (e.g. \citealt{jimenesgarate2002,fuerst2013}). We do find some differences in the blackbody temperature and normalization between Observations H1 and H4, which is likely due to a combination of the short pre-dip exposure time for Observation H4 and small changes in the spectral continuum shape with superorbital phase (e.g., \citealt{fuerst2013}).

We see significantly different spectral shapes in Observations H2 and H3 than we do in H1 and H4. While some spectral differences with superorbital phase are to be expected, the low flux and lack of strong pulsations during Observations H2 and H3 indicate that these observations are somewhat unusual for Her X-1. We suggest that the neutron star and central accretion region were obscured during these observations because of the reduced flux and lack of pulsations.

Using the same warped disk model from B20, we were able to simulate a simple warped disk irradiated by either a pencil or a fan beam emission geometry and calculate the simulated pulse profiles that would be observed for different precession angles of the disk. We used the previous disk modeling of \cite{scott2000} and \cite{leahy2002} to guide our choice of disk geometry. We ultimately found that both the pencil and fan beam emission models, which were fit independently, were capable of reproducing the observed pulse profiles and that the simulated disk precession phase was in good agreement with the superorbital phase of our observations. However, the pencil beam emission geometry provides a slightly better fit to the observed pulse profiles, which can be seen by the smaller values of $r$ for each observation in Figures \ref{fig:hdisksimpppencil} and \ref{fig:hdisksimppfan}. However, we would like to note that the geometries used in this model are simplistic for the purpose of highlighting the contribution of the precessing disk. It is likely that the accretion geometry of Her X-1 is more complex than this model suggests.

In both the fan and pencil beam emission geometries we find that the preferred beam geometry is strongly non-antipolar. We demonstrate this conclusion by showing a simulated pulse profile from antipodal pencil beams in Figure \ref{fig:highrespp}. The shape of the simulated pulse profile from antipodal beams is not a good fit to our observed pulse profiles. \cite{kraus1995} identified distortions in the dipolar field of neutron stars as a possible cause of asymmetry in pulse profiles. \cite{blum2000} found that the energy-resolved pulse profiles from Her X-1 suggested a slightly distorted dipolar field. B20 and \cite{hickoxvrtilek2005} also found this preference for non-antipolar beams. This preference may suggest that the structure of magnetized accretion flows are more complex than the current scope of our warped disk model. Future modeling efforts would benefit from considering more complex emission geometries (e.g., \citealt{koliopanos2018, iwakiri2019}), physically motivated accretion column models (e.g., \citealt{sokolovaprep}), or the effects of light bending from the neutron star (e.g., \citealt{falknersuba}; \citealt{falknersubb}).

Observation H4 contains a pre-eclipse dip from Her X-1 with both \xmm\ and \nustar. Examining the pulsed fractions (Figure \ref{fig:herlc}) and energy-resolved light curves (Figure \ref{fig:hdiplc}) both show strong absorption of the soft X-ray emission consistent with obscuration by part of the accretion disk. The data do not show evidence of a periodic spiking signal previously seen by \cite{vrtilek1985}.

\section{Conclusion}
In this work we performed a broad-band X-ray timing analysis of Her X-1 during its 35 day superorbital cycle. Our series of four joint \xmm\ and \nustar\ observations sampled a single superorbital cycle; however, we focus on the first and fourth observations in this series which had sufficient signal to noise to create energy-resolved pulse profiles in narrow energy bandpasses. We supplemented our missing coverage of the superorbital phase with an archival \xmm\ observation at $\phi_{\text{SO}}$=0.60. We found that the soft ($<1$ keV) and hard ($>8$ keV) pulse profiles had similar shapes and relative phase offsets in Observations H1 and H4, which we expected from a warped, precessing accretion disk that has returned to its original position. The joint spectral fits also showed periodicity with superorbital phase, which supports the precessing disk scenario. We use the simple warped disk model used by B20 to simulate our observed pulse profiles and find that they are consistent with reflection off of a precessing disk. We find a strong preference for non-antipolar neutron star beam geometry, which is consistent with the results of B20. Updates to this model with more physically motivated beam geometries could further test this non-antipolar preference. We also examined the energy resolved light curves of the pre-eclipse dip seen in Observation H4. We find strong absorption of the soft X-rays and variability in the hard X-rays consistent with previous observations. We do not see evidence of a periodic signal within the dip.

\acknowledgements
We would like to thank the anonymous referee for comments and suggestions that greatly improved the clarity and substance of this paper. We thank Lorenzo Ducci and Matteo Bachetti for helpful discussions relating to the spectral and timing analyses, respectively. We would like to thank the \textit{NuSTAR} Galactic Binaries Science Team for comments and contributions. MCB acknowledges support from NASA grant numbers NNX15AV32G and NNX15AH79H. This research made use of NuSTARDAS, developed by ASDC (Italy) and Caltech (USA), XMM-SAS developed by ESA, ISIS functions (ISISscripts) provided by ECAP/Remeis observatory and MIT (http://www.sternwarte.uni-erlangen.de/isis/), and the MAXI data provided by RIKEN, JAXA and the MAXI team.

\software{HEAsoft (v6.26.1; HEASARC 2014), NuSTARDAS, SAS (\citealt{xmmsas}), Stingray (\citealt{stingray}), Xspec (v12.10.1; \citealt{arnaud1996}), MaLTPyNT (\citealt{hendrics})}

\appendix

\section{Extended Timing Analysis}
\begin{figure*}
\centering
\includegraphics[scale=0.8]{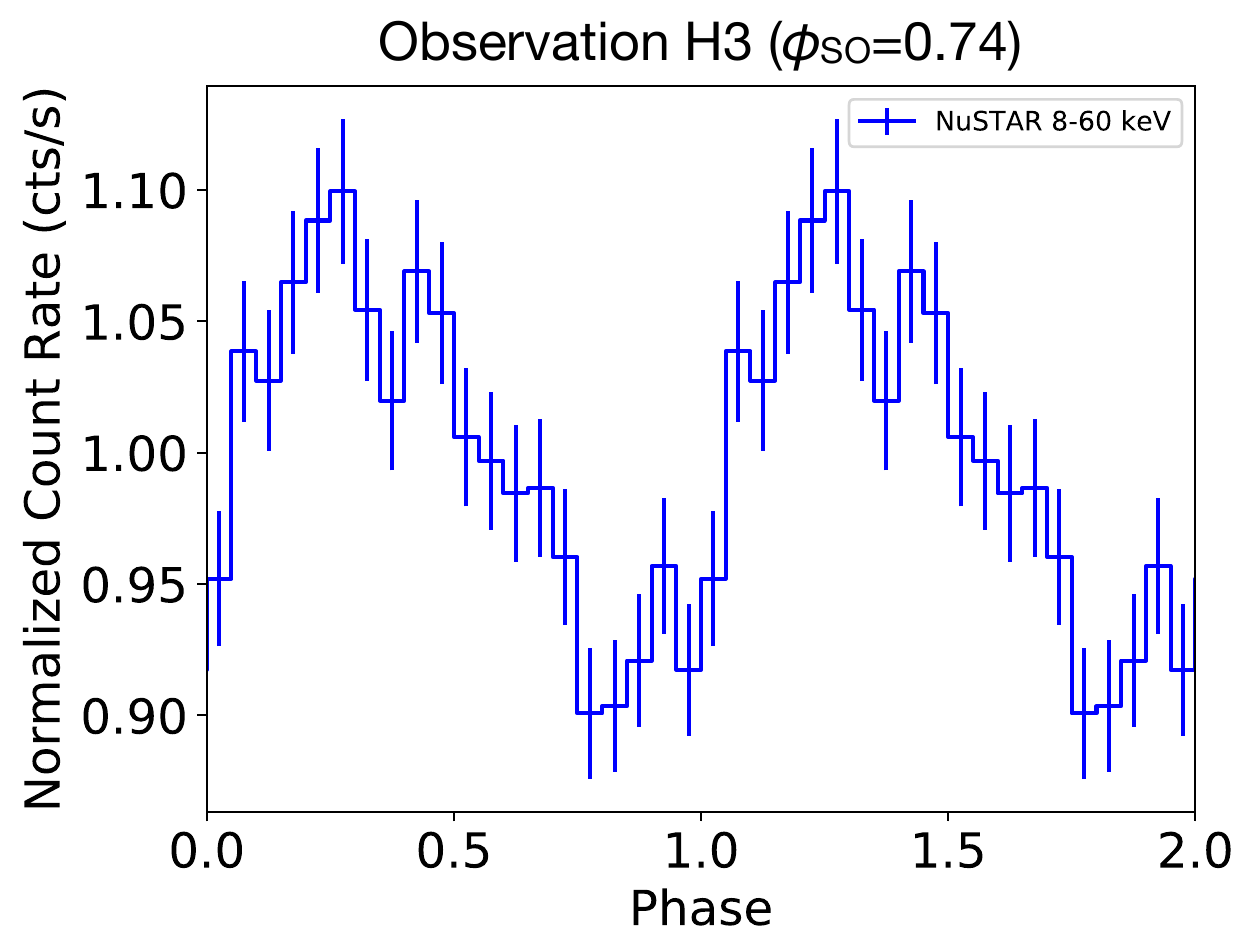}
\caption{The \nustar\ 8--60 keV pulse profile from the first 5 ks of Observation H3. The pulses in this observation are weaker and have a different pulse shape than those seen in Observations H1 and H4. For clarity, we have shifted this pulse profile so that the peak aligns in phase with the hard peak from Observation H1.}
\label{fig:h3pp}
\end{figure*}

Although we could not use our warped disk model on Observation H3 due to the lack of \xmm\ pulsations, we were able to extract a \nustar\ 8--60 keV pulse profile. To create this pulse profile, we selected the first 5 ks of \nustar\ data which were most strongly pulsed and followed the method described in Section \ref{Hdisksec:analysis}. We show the \nustar\ pulse profile of Observation H3 in Figure \ref{fig:h3pp}. The pulse profile shows a single broad peak that is atypical for hard pulses from Her X-1.

When selecting the energy bands to be used in our hard and soft pulse profiles, we found it necessary to examine the energy dependence of the pulse profile to make an appropriate selection. This decision was motivated heavily by \cite{ramsay2002}, who examined the energy dependence of their \xmm\ observations and found changes in the soft pulse profile beginning around 0.8 keV (see Figure 2 in \cite{ramsay2002}. We created similar figures by filtering the \xmm\ data for Observations H1 and H4 into the following energy bins: 0.3--0.7 keV, 0.8--1.2 keV, 1.5--3 keV, 3--6 keV, 6.2-6.6 keV, and 7--12 keV. We also filtered the \nustar\ data for Observations H1 and H4 into energy bins consisting of 3--6 keV, 6.2--6.6 keV, 7-12 keV, 12.4-30 keV, 30.4--60 keV. We based these energy bins on those used by \cite{ramsay2002}, but adjusted the energy ranges slightly to suit joint \xmm\ and \nustar\ observations. We also filtered the pulsed portion of the \nustar\ data from Observation H3 into the same energy bands used for the \nustar\ data of Observations H1 and H4. As mentioned in Section \ref{Hdisksec:obs}, we were unable to extract coherent energy-resolved pulse profiles from the \xmm\ data of Observation H3.

\begin{figure*}
\includegraphics[scale=0.67]{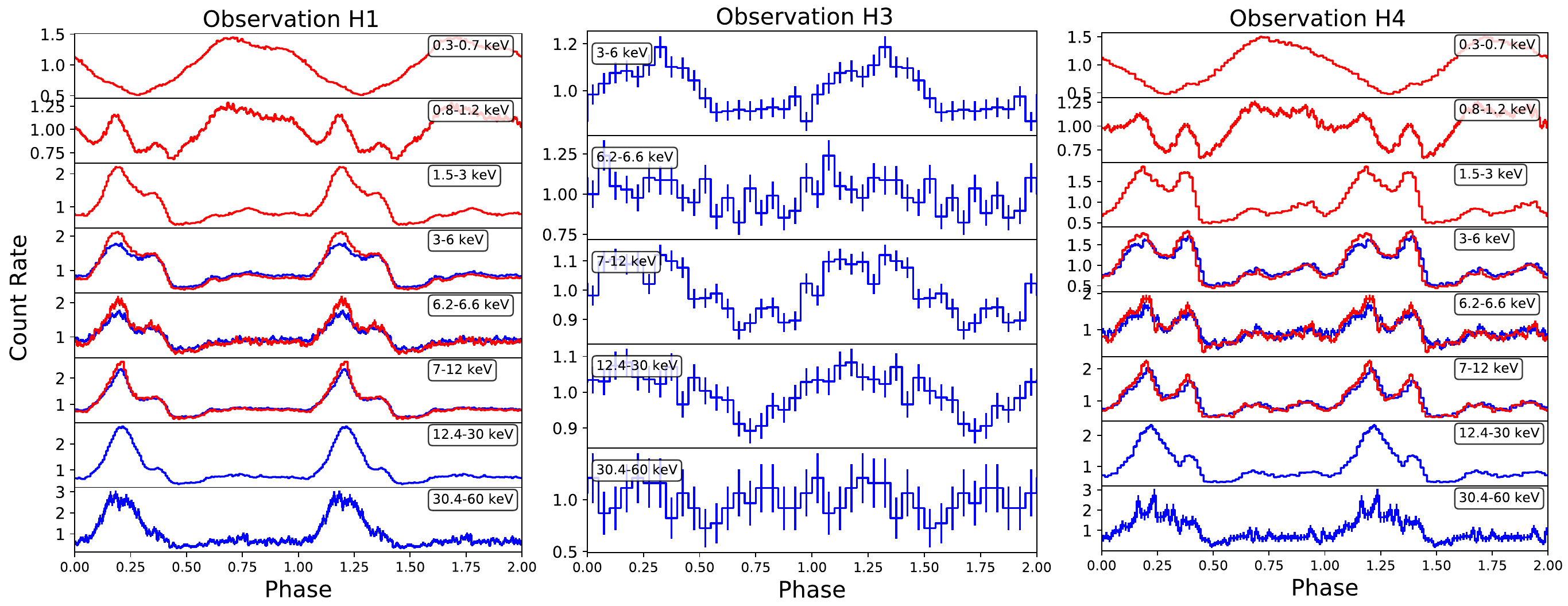}
\caption{Energy resolved pulse profiles for Observations H1, H3, and H4. Red profiles are \xmm\ and blue profiles are \nustar. In order to show the pulse profiles in detail while maintaining high signal-to-noise, we varied the resolution with which we produced these profiles to match the effective exposure of the pulsed emission: Observation H1 profiles contain 128 phase bins, Observation H3 profiles contain 20 phase bins, and Observation H4 profiles contain 70 phase bins. We have also shifted the profiles of Observations H3 and H4 so that the hard pulse peak aligns in phase with the peak from H1, for clarity. We note that there is strong energy dependence in the soft band, which is illustrated by the emergence of the primairy pulse peak as early as 0.8 keV. The softest energy band (0.3--0.7 keV) shows a smooth, single peaked profile that we expect from reprocessed emission (e.g., \citealt{hickox2004}). The 0.8--1.2 keV band appears to be a mix of the reprocessed emission and the harder, sharply pulsed profiles that dominates at energies above 1.5 keV. In Observation H3 we are unable to produce energy resolved pulse profiles from the \xmm\ data.}
\label{fig:enrespp}
\end{figure*}

We folded the energy-resolved data by the best fit period for the corresponding observation. We varied the resolution with which we plotted the pulse profiles to match the effective exposure of the pulsed emission from each observation: Observation H1 profiles contain 128 phase bins, Observation H3 profiles contain 20 phase bins, and Observation H4 profiles contain 70 phase bins. We have also shifted the profiles of Observations H3 and H4 so that the hard pulse peak aligns in phase with the peak from H1. We show the resulting pulse profiles in Figure \ref{fig:enrespp}.

Figure \ref{fig:enrespp} shows that the soft pulse profiles are highly energy dependent and that the sharp, notched peak that defines the hard pulse profile begins to emerge around 0.8 keV. In order to isolate the soft, reprocessed emission, we therefore selected the energy range of 0.3--0.7 keV for our soft energy band in this work.

While the pulse profiles of Observations H1 and H4, and their energy dependence, are almost identical (as we expect from the precessing disk scenario), the pulse profiles from Observation H3 differ significantly. The pulsations are generally weaker and the pulse is much broader than the pulses in Observations H1 and H4, and lacks the distinctive notch. Some differences in pulse shape can be expected from the processing disk scenario, in which different parts of the accretion column are visible at this superorbital phase.

Interestingly, we note that the energy resolved pulse profiles shown by \cite{ramsay2002} are significantly different in shape than those from our series, despite the similar energy bins used. Some of these changes may be expected from the differences in superorbital phase (\cite{ramsay2002} observations fall at superorbital phases 0.17, 0.26, and 0.60, compared to the phases of Observations H1, H3, and H4 of 0.20, 0.74, and 1.14). However, the magnitude of these differences imply that the pulse shape has changed between these two series. The pulse profiles shown in this work show more similarity to the pulse profiles presented in \cite{staubert2009}.

\pagebreak
\bibliography{my_bib.bib}

\end{document}